%% file: main.tex
\documentclass[fleqn,usenatbib]{mnras}

\usepackage{newtxtext,newtxmath}

\usepackage[T1]{fontenc}

\DeclareRobustCommand{\VAN}[3]{#2}
\let\VANthebibliography\thebibliography
\def\thebibliography{\DeclareRobustCommand{\VAN}[3]{##3}\VANthebibliography}

\usepackage{graphicx}	
\usepackage{amsmath}	
\usepackage{siunitx}
\usepackage{booktabs}
\usepackage{CJKutf8}
\usepackage{bm}
\usepackage{pdflscape}
\usepackage{caption}
\title[Metallicity in PCFNet Protoclusters at $z \sim 3.4$]{Diverse Metallicities in Protoclusters at $z\sim3.4$ detected by Deep Learning}
\author[Y. Takeda et al.]{
Yoshihiro Takeda,$^{1}$\thanks{E-mail: y.takeda.astron@gmail.com}
Nobunari Kashikawa,$^{1,2}$
Kei Ito,$^{3,4}$
Jun Toshikawa,$^{5,6}$
Mariko Kubo,$^{6,7}$
\newauthor
Rieko Momose,$^{8,9}$
Yongming Liang,$^{9,10}$
Takehiro Yoshioka,$^{1}$
Junya Arita,$^{1}$
Hiroki Hoshi,$^{1}$
\newauthor
Shunta Shimizu,$^{1}$
Hisakazu Uchiyama,$^{9,11}$
Satoshi Kikuta,$^{1}$
Ryo Emori$^{1}$
and 
Kentaro Koretomo $^{1}$
\\
$^{1}$Department of Astronomy, School of Science, The University of Tokyo, 7-3-1 Hongo, Bunkyo-ku, Tokyo, 113-0033, Japan\\
$^{2}$Center for the Early Universe, The University of Tokyo, 7-3-1 Hongo, Bunkyo-ku, Tokyo, 113-0033, Japan\\
$^{3}$ Cosmic Dawn Center (DAWN), Copenhagen, Denmark\\
$^{4}$ DTU Space, Technical University of Denmark, Elektrovej 327, DK2800 Kgs. Lyngby, Denmark\\
$^{5}$ Nishi-Harima Astronomical Observatory, Center for Astronomy, University of Hyogo, Sayo, Hyogo 679-5313, Japan\\
$^{6}$Astronomical Institute, Tohoku University, Aoba-ku, Sendai 980-8578, Japan\\
$^{7}$Department of Physics and Astronomy, School of Science, Kwansei Gakuin University, 1 Gakuen Uegahara, Sanda, Hyogo 669-1330, Japan\\
$^{8}$Observatories of the Carnegie Institution for Science, 813 Santa Barbara Street, Pasadena, CA 91101, USA\\
$^{9}$National Astronomical Observatory of Japan, 2-21-1 Osawa, Mitaka, Tokyo 181-8588, Japan\\
$^{10}$Institute for Cosmic Ray Research, The University of Tokyo, Kashiwa, Chiba 277-8582, Japan\\
$^{11}$Department of Advanced Sciences, Faculty of Science and Engineering, Hosei University, 3-7-2 Kajino-cho, Koganei, Tokyo 184-8584, Japan
}

\date{Accepted XXX. Received YYY; in original form ZZZ}

\pubyear{\the\year{}}

\begin{document}
\label{firstpage}
\pagerange{\pageref{firstpage}--\pageref{lastpage}}
\maketitle

\begin{abstract}
We present spectroscopic follow-up observations with Subaru/MOIRCS of four protocluster candidate regions identified by the deep-learning model PCFNet.
In three candidate regions, we confirm four protoclusters at $z=3.316,\ 3.403,\ 3.408\ \mathrm{and}\ 3.425$, exhibiting galaxy density peaks over five times the uniform distribution.
The detection rate of confirmed members aligns with PCFNet's predictions, underscoring its effectiveness.
Analysis of spectral energy distributions (SEDs), incorporating up to ten photometric bands $(uu^*grizyJHK)$, reveals that protocluster members lie on the star-formation main sequence.
The masses of the most massive halos in the protoclusters are estimated from the most massive stellar mass of the member galaxies, yielding $\log M_\mathrm{halo}/M_\odot=12.4\text{--}12.8$, which are consistent with those 
at this redshift predicted by cosmological simulations and are comparable to those of the progenitors of typical local clusters ($M_\mathrm{halo}^{z=0}\sim10^{14}M_\odot$).
Using the Voronoi Tessellation Monte Carlo technique based on the probability of protocluster membership from PCFNet, we find that our protoclusters exhibit a significantly higher SFRD than the cosmic SFRD at $z\sim3.4$.
Based on the [O\textsc{iii}]/H$\beta$ emission line ratio, two protoclusters show marginally ($1.4\text{--}2\sigma$) higher average metallicities than the field, while the other protoclusters are statistically consistent with the field.
These diverse metallicities highlight the environmental impact within protoclusters detected by PCFNet, likely reflecting the complex opposing interplay between metal-rich gas recycling and metal-poor gas inflow, which may depend on the surrounding environment of each protocluster.
This work demonstrates PCFNet's capability to unveil diverse samples of high-redshift protoclusters, paving the way for a more comprehensive understanding of their role in galaxy evolution.
\end{abstract}




\section{Introduction}
In the local Universe, it is well established that galaxy properties such as morphology, stellar mass, and star-formation activity correlate with environment: dense regions host a larger fraction of early-type galaxies, while low-density regions contain a higher fraction of star-forming spirals \citep{Dressler1980,Postman1984}.
The characteristic mass of galaxies increases with local density \citep{Baldry2006}, and the fraction of high-mass galaxies is higher in clusters than in the field \citep{Etherington2017}.
The star formation rate (SFR) also shows a strong dependence on environment, with galaxies in dense regions exhibiting systematically lower specific star formation rates and a higher quenched fraction compared to those in low-density environments \citep[e.g.,][]{Lewis2002,Kauffmann2004}.
According to theoretical models of structure formation, galaxy formation proceeds more rapidly in overdense regions at high redshift \citep{Cen2000, Springel2006}.
These environmental effects have attracted significant interest in recent years, with key questions focusing on how and when such effects emerge in overdense environments.
It is one of the remaining mysteries of galaxy formation and evolution.

Protoclusters, high-redshift structures destined to evolve into clusters with $M_\mathrm{halo}^{z=0}>10^{14}\,M_\odot$ \citep{Overzier2016}, are ideal laboratories for investigating environmental effects, which are actively manifesting during their formation and evolution \citep{Chiang2013}.
Over the past few decades, various protocluster search techniques have been developed and applied, leading to the discovery of protoclusters \citep[e.g.][]{Cucciati2014,Lemaux2014,Miller2018,Oteo2018,Staab2024,Venemans2007,Jiang2018,Hu2021}.
Besides, the advent of the James Webb Space Telescope (JWST) has greatly accelerated the identification of high redshift overdensities beyond $z \sim 5\text{--}6$ \citep{Morishita2023, Hashimoto2023, Helton2024, Fudamoto2025}; they are often the most overdense structures in their redshifts, and many of them are thought to be the ancestors of massive clusters like the Coma cluster with $M_\mathrm{halo}^{z=0}>10^{15}M_\odot$.
Such most massive protoclusters are expected to be located at nodes in the large-scale structure \citep{Yajima2022}, where cold inflow is expected to occur due to their deep gravitational potentials \citep{Dekel2006}.
While studying such extreme environments is important, the evolutionary paths of protoclusters are diverse \citep{Muldrew2015,Remus2023}, and focusing solely on the most massive protoclusters at each epoch is insufficient to understand the general picture of protocluster and galaxy evolution.

However, it is challenging to extend the protocluster sample toward the low-mass regime, especially at $z>3$.
The main issue arises from the fact that commonly used protocluster search techniques detect pronounced overdensities on the sky, which can lead to a bias towards detecting extremely massive protoclusters while potentially missing smaller ones. 
In recent years, the number of observed moderate protoclusters at high redshift has been increasing by using high-quality photometric redshift data combining deep multi-wavelength survey observations \citep{Hung2025, Newman2025, Toni2025}. 
However, it is difficult to use the same method in wide-field surveys such as the Hyper Suprime-Cam Subaru Strategic Program \citep[HSC-SSP; ][]{Aihara2018, Aihara2022} and the Legacy Survey of Space and Time \citep[LSST; ][]{Ivezic2019}.

Furthermore, the metallicity measurements in these early structures are crucial to understanding the mechanisms of recycling of enriched gas and pristine gas inflow in overdense regions. 
The gas-phase metal abundances in protoclusters at $z\sim2$ have yielded diverse results, with some studies indicating enriched environments compared to the field \citep[e.g.]{Shimakawa2015,Kulas2013}, while others find no significant difference or even lower metallicities \citep[e.g.][]{Valentino2015,Kacprzak2015}. 
These discrepancies might stem from the inherent diversity among protoclusters themselves, such as variations in their evolutionary stage, mass, or gas accretion history, highlighting the need for larger, more representative samples in a broad range of environments to clarify the processes driving chemical evolution in high-redshift overdensities.

To overcome the above difficulties, a deep-learning-based protocluster detection method, called PCFNet \citep{Takeda2024}, has been developed. 
PCFNet outputs the probability that galaxies are in a protocluster using only optical broad-band photometry, taking into account not only their sky distribution but also the line-of-sight distance inferred from their magnitudes and colours.
PCFNet is optimised for HSC-SSP, one of the widest optical ($grizy$-band) surveys, and $121$ protocluster candidates at $z\sim3.8$ are detected from $17\mathrm{\,deg^2}$ of the DEEP layer of HSC-SSP.

The paper reports the spectroscopic follow-up observations of four protocluster candidate regions detected by PCFNet.
This study aims to verify the reliability of PCFNet by conducting spectroscopic observations of protoclusters detected by the method, and to investigate the characteristics of these protoclusters at $z\sim3.8$, thereby improving our understanding of galaxy evolution in typical protocluster environments at this cosmic epoch.
Section~\ref{sec:method} describes the observation and the analysis,
Section~\ref{sec:results} presents the results of the spectroscopic identification and the physical properties of the protoclusters, and Section~\ref{sec:discussion} discusses the implications of these findings.
Finally, we conclude our work in Section~\ref{sec:conclusion}.
We assume the following cosmological parameters estimated by the Planck mission results \citep{PlanckCollaboration2020} unless otherwise noted: $\Omega_m = 0.30966$, $\Omega_\Lambda = 0.69034$, and H$_0$ = 67.66 km s$^{-1}$ Mpc$^{-1}$, and we use the AB magnitude system \citep{Oke1983}.

\section{Method}
\label{sec:method}
\subsection{Protocluster candidate catalogue by PCFNet}

We construct a protocluster candidate catalogue using the deep learning framework PCFNet \citep{Takeda2024}.
The motivation for developing such a catalogue is to efficiently locate protoclusters at $z>3$, where traditional surface-density-based methods often fail due to projection effects and limited sensitivity to low-mass systems.
A brief catalogue description is provided here; readers are referred to \citet{Takeda2024} for further detailed explanations.

For training and validation, PCFNet is trained on a realistic mock galaxy catalogue, PCcone \citep{Araya-Araya2021}, which is generated from large-volume cosmological simulations designed to reproduce the photometric and spatial distributions of photo-$z$ selected galaxies observed in the HSC-SSP and LSST surveys.
This mock reproduces key observational properties such as magnitude limits, photometric uncertainties, and selection completeness, ensuring that the network learns to capture environmental features comparable to the real data.
We set the observing conditions of PCcone to match those of the HSC-SSP Deep and UltraDeep layers and use $g$-dropout galaxies selected from the mock catalogue, which effectively trace the large-scale structure at $z\sim3.8$ \citep{Ono2018}.
The selection criteria for $g$-dropout galaxies are as follows:
\begin{equation}
     \begin{split}
     &(g-r > 1.0)\\
     \land\quad &(-1.0 < r-i < 1.0)\\
     \land\quad &(1.5(r-i) < g - r -0.8)\label{eq:gdropout}.
     \end{split}
\end{equation}
We check the magnitude distribution of the selected $g$-dropout galaxies in PCcone and confirm that it is consistent with that of the actual HSC-SSP data down to $i=26$ mag.
PCFNet is trained with the mock $g$-dropout catalogue with the limiting magnitude of $i=26$ mag, using as input the sky coordinates and $g, r, i$-band photometry of galaxies within a $5'$ radius of each target.
PCFNet outputs the probability of each galaxy being a protocluster member, expressed as a significance, $\sigma_\mathrm{prob}$.
Galaxies with $\sigma_\mathrm{prob}>2.5$ are defined as protocluster member candidates, and protocluster candidates are identified by grouping these galaxies using the \texttt{findpeaks} algorithm \citep{Taskesen2020}.
Compared to the conventional 2D surface number density approach, PCFNet identifies about five times more protocluster member galaxies at $z\sim3.8$ with higher accuracy, and detects 24 times more protoclusters with typical halo masses ($M_{\mathrm{halo}}^{z=0}<10^{14.5}M_\odot$).

The model is then applied to the HSC-SSP S20A DEEP/UltraDEEP data \citep{Aihara2022}, which provides deep optical imaging with $5\sigma$ depth magnitude of $g=27.4$, $r=27.0$, and $i=26.8$ mag and the average seeing of $0.66''$--$0.83''$ in the DEEP layer.
Note that the HSC-SSP Deep field in the COSMOS region, referred to as E-COSMOS in the HSC-SSP project, extends beyond the footprint of the original COSMOS field. 
We use the convolved point-spread function (PSF) photometries measured with the same position of each band (\texttt{forced}) as the object magnitudes.
The objects are selected with $r<m_{\mathrm{lim,5}\sigma}$, $23<i<26$ mag, and good quality flag \citep[see for more details, ][]{Takeda2024} to ensure reliable photometry and low contamination from foreground objects, where $m_{\mathrm{lim,5}\sigma}$ is the $5\sigma$ depth magnitude.
We select $g$-dropout galaxies from the HSC-SSP photometric catalogue using the criteria in Equation~\ref{eq:gdropout} and apply PCFNet to this sample, resulting in the identification of 121 protocluster candidates over an area of approximately 17 deg$^2$.

\subsection{Target selection}
\label{sec:sample}
Our target selection consisted of two steps:
(i) selection of protocluster candidate targets for follow-up observations with MOIRCS, and
(ii) selection of such candidates to which slits would be assigned.
We describe these two steps separately below.

\subsubsection{Selection of protocluster candidate targets}
The $g$-dropout galaxies are typically found in the redshift range of $3.3<z<4.5$ \citep{Ono2018}. 
Ground-based observations of the [O\textsc{iii}] $\lambda5007$ line become increasingly difficult at $z \gtrsim 3.8$ because the line is shifted to $\gtrsim 2.3 \micron$, where the thermal background becomes severe.
Although ground-based spectroscopic observations are still possible longward of the $K$ band for very bright sources, observations for typical high-redshift star-forming galaxies in the wavelengths are generally difficult.
We therefore restricted our follow-up to candidate regions likely to lie at $3.3 < z < 3.8$ using photometric redshifts from a joint multi-wavelength catalogue.
We used the multi-wavelength ($uu^*grizyJHK$) joint catalogue, which is composed of wide survey collaborations: HSC-SSP, the CFHT Large Area $u$-band Deep Survey \citep[CLAUDS,][]{Sawicki2019}, and the Deep UKIRT Near-Infrared Steward Survey (DUNES$^2$, Egami et al., in prep.). 
The joint catalogue provides photometric redshifts estimated with Mizuki \citep{Tanaka2015,Tanaka2018}.
The parameter \texttt{photoz\_risk\_best} in the catalogue quantifies the probability of an outlier, which is defined as a source satisfying $|z_\mathrm{phot} - z_\mathrm{true}| > 0.15(1 + z_\mathrm{true})$, where $z_\mathrm{true}$ denotes the reference redshift used in the calibration of the photometric-redshift performance in the catalogue construction.
For sources with $i < 26.5$ mag and \texttt{photoz\_risk\_best} $< 0.5$, the median photometric-redshift uncertainty at $3.3 < z_\mathrm{phot} < 3.8$ is approximately $(z_\mathrm{68,max} - z_\mathrm{68,min})/(1 + z_\mathrm{phot}) \simeq 0.15$, where $z_\mathrm{68,max}$ and $z_\mathrm{68,min}$ are the upper and lower bounds of the 68\% confidence interval.
It is important to note that while PCFNet is primarily designed for ultra-wide surveys to identify protocluster candidates from photometric data alone, in this study, we also used photometric redshifts to ensure successful follow-up spectroscopic observations.

We selected targets from the 121 protocluster candidates identified by PCFNet \citep{Takeda2024} through the sequence summarised in Table~\ref{tab:protocluster_selection}. 
We defined galaxies with $\sigma_\mathrm{prob} > 2.5$ as high-probability protocluster member candidates, following Takeda et al. (2024).
We first selected regions containing more than ten such member candidates in the MOIRCS \citep{Suzuki2008, Ichikawa2006} FoV of $4 \times 7\ \mathrm{arcmin}^2$.
We then required that the candidate contain at least five member candidates in the target redshift interval, $3.4 < z_\mathrm{phot} < 3.8$.

To favour regions in which the [O\textsc{iii}] $\lambda5007$ line is accessible with MOIRCS, we used $g$-dropout galaxies in the MOIRCS FoV with reliable photometric redshifts (\texttt{photoz\_risk\_best} $<0.3$) and $z_\mathrm{phot}\geq3.4$, and retained regions with a mean photometric redshift of $\langle z_\mathrm{phot}\rangle\leq3.8$.
Finally, we visually inspected the remaining candidates and excluded regions affected by scattered light or nearby bright stars, already scheduled for observations by other projects, or lacking a central concentration of protocluster member candidates within the MOIRCS FoV. 
This selection yielded eight candidate regions.

\begin{table*}
    \centering
    \caption{Selection of protocluster candidate targets for MOIRCS follow-up. The number in the final column is the number of candidates remaining after each selection step.}
    \label{tab:protocluster_selection}
    \begin{tabular}{p{0.17\textwidth}p{0.62\textwidth}r}
        \hline
        Step & Criterion & Candidates remaining \\
        \hline
        Initial sample & PCFNet protocluster candidates & 121 \\
        Member richness & $N_{\mathrm{mem}}(\sigma_\mathrm{prob}>2.5)>10$ in the MOIRCS FoV & 37 \\
        Member redshift & $N_{\mathrm{mem}}(3.4<z_\mathrm{phot}<3.8)\geq5$ & 33 \\
        FoV redshift & $\langle z_\mathrm{phot}\rangle\leq3.8$ for $g$-dropouts with $z_\mathrm{phot}\geq3.4$ and \texttt{photoz\_risk\_best}$<0.3$ & 21 \\
        Visual inspection & Exclude scattered light/nearby bright stars (4), other projects (4), or non-central member distributions (5) & 8 \\
        \hline
    \end{tabular}
\end{table*}

From these, we selected the four highest-priority regions that were feasible in the S24B semester.
These regions contain 11--47 member candidates, which have a high protocluster membership probability predicted by PCFNet.
Three of the four target regions were originally identified by PCFNet, while the DEEP2-3\_ID2 region had also been reported previously as a photometrically selected protocluster candidate by \citet{Toshikawa2024}. 
To our knowledge, however, no spectroscopic confirmation of that candidate had been presented prior to this work. 
Our COSMOS\_ID0 target lies in E-COSMOS, outside the original COSMOS footprint, where ancillary multi-wavelength coverage and previous protocluster searches are less extensive. We find no clear association with previously reported overdensities, noting these differences in coverage and search efforts.
We also find no obvious previous report of a comparable overdensity at the position of ELAIS-N1\_ID4.

\subsubsection{Selection of galaxy targets to which slits would be assigned}
After selecting the protocluster candidate targets, we prioritised slit assignment according to the PCFNet membership significance.
Targets with $\sigma_\mathrm{prob} > 2.5$ were given the highest priority, followed by those with $1.5 < \sigma_\mathrm{prob} \le 2.5$.
Remaining slits were assigned to galaxies with $\sigma_\mathrm{prob} \le 1.5$, for which we additionally required $3.3 < z_\mathrm{phot} < 3.9$ to account for photometric-redshift uncertainties.
As a result, the slit-allocation completeness, defined as the fraction of targets with $\sigma_\mathrm{prob} > 2.5$ that were allocated slits, was 55\% (6/11) in ELAIS-N1\_ID4, 43\% (20/47) in DEEP2-3\_ID1, 37\% (11/30) in DEEP2-3\_ID2, and 47\% (16/34) in COSMOS\_ID0.
The remaining science slits in Table \ref{tab:obssummary} were allocated to targets with $\sigma_\mathrm{prob} \leq 2.5$.

\subsection{MOIRCS Observation}
\label{ssec:moircs}

Observations were conducted with MOIRCS over three nights: in August 2024, September 2024, and January 2025.
We utilised the VB-K grism (Ebizuka et al., in preparation) and the OC1--3 filter, covering the wavelength range of $\lambda\sim1.9\text{--}2.35\,\mathrm{\mu m}$ ($\sim4400\text{--}5300$\AA~in rest-frame at $z=3.4$). This range includes the [O\textsc{iii}]$\lambda5007$ emission line for galaxies at $2.8<z<3.7$.
The slits have a $0.8''$ width, providing a spectral resolution of $R \sim 1600$.
Details of each observation are summarised in Table~\ref{tab:obssummary}.
The total number of science targets is 115, and a separate mask was used for each target region.
The overall observations were conducted in good condition, but the observing run in August 2024 suffered from poor seeing conditions ($\sim1''\text{--}2''$) due to strong winds and had a short exposure time as it was conducted as a gap-filler programme. Consequently, these data were excluded from the subsequent analysis.
We took each shot with a five-minute exposure time and dithered along the slits by 2.4--2.6$''$, which is shorter than the minimum slit length of $7.5''$.
The mean total integration time was approximately $130$ minutes, achieving a signal-to-noise ratio (S/N) of $\sim3$ for a flux density of $f_\lambda \sim 2\times10^{-18} \mathrm{erg\,s^{-1}\,cm^{-2}\,\mbox{\AA}^{-1}}$.

Data reduction was performed using the MCSMDP \citep{Yoshikawa2010}, an \texttt{iraf}-based MOIRCS MOS data reduction package.
The wavelength calibration was performed using sky lines, followed by sky subtraction between dithered frames, and then median combining the results.
The flux was calibrated by telluric standard stars (HIP12858 and HIP56147), which were observed with the same slit in the object masks on each observing night.

\begin{table*}
	\centering
	\caption{MOIRCS Observation details.}
	\label{tab:obssummary}
	\begin{tabular}{lccccl}
		\toprule
		Region & Date & Exp time & \# of science slits & comments\\
		 &  & [min] & & &\\
		\midrule
		 ELAIS-N1\_ID4 & 2024 Aug 21 &30 & 27 & bad seeing ($\sim1''\text{--}2''$)\\
		  & 2024 Sep 13 & 100 &  &\\
		DEEP2-3\_ID1 & 2024 Sep 13 & 130 & 30 &\\
		DEEP2-3\_ID2 & 2024 Sep 13 & 150 & 29 &\\
		COSMOS\_ID0 & 2025 Jan 18 & 250 & 29 &\\
		\bottomrule
	\end{tabular}
\end{table*}

\subsection{Line identification}
\label{ssec:line_id}

We visually inspected the two-dimensional spectra to identify emission lines and determined the positions of the objects based on the profiles of the emission lines collapsed along the wavelength direction, from which we extracted one-dimensional spectra.
Emission lines were simultaneously fitted with three Gaussian profiles for H$\beta$, [O\textsc{iii}]$\lambda4959$ and [O\textsc{iii}]$\lambda5007$ with fixed wavelength ratios.
The line flux is measured by integrating the best-fit Gaussian profile.
The flux uncertainty is derived from the parameter covariance matrix of a weighted fit, where the weights are set by a Poisson-based variance frame computed from the sky-unsubtracted counts. The $1\sigma$ uncertainty is estimated by propagating from this covariance.
A single emission line was considered to be primarily either the [O\textsc{iii}]$\lambda5007$ at $z\sim3.5$ or the Pa$\alpha$ at $z\sim0.2$; these can be distinguished based on whether the $g$-band flux is detected. 
Other possible identifications, such as H$\alpha$ or [O\textsc{ii}]$\lambda3726,3729$, were further assessed using the multi-band SED fitting described in Section \ref{ssec:sed_fitting}.

We detect emission lines from 60 targets across the four observed regions at $\geq3\sigma$ significance in the integral line flux, of which 40 exhibit multiple emission features, while the remaining 20 show only a single detected line.
Three out of the 20 single-line objects are identified as Pa$\alpha$ emitters at $z\sim0.3$ through their detections in the $u$ band, while the remaining objects are identified as [O\textsc{iii}]$\lambda5007$ emitters based on their SED fitting results.
All 1D and 2D spectra are presented in Fig.~\ref{fig:all_2d_1d}, and the properties of the emission detected objects are summarised in Table~\ref{tab:obsemission}.
The 55 objects, which did not exhibit emission lines, are considered to have either emission lines that are fainter than the detection limits or no emission lines in the $K$-band wavelength range (see Appendix~\ref{sec:detect_nondetect}).

\subsection{SED fitting}
\label{ssec:sed_fitting} 
We employ the SED fitting code \textsc{Bagpipes} \citep{Carnall2018} to derive the physical properties of the spectroscopically identified galaxies from their multi-band photometry and spectroscopic redshifts.
\citet{Bruzual2003} Stellar Population Synthesis (SPS) model with Kroupa initial mass function \citep[IMF, ][]{Kroupa2001}, the analytic IGM attenuation model of \citet{Inoue2014} and the Calzetti dust attenuation law \citep{Calzetti2000} are assumed.
We adopt a delayed star formation history (SFH) model, expressed as $\mathrm{SFR}\propto t \exp(-t/\tau)$, where $t$ represents the age of the galaxy and $\tau$ is the characteristic timescale for the decline in star formation.
The redshift is fixed at spec-$z$ for spectroscopically identified galaxies, if detected, otherwise we set the redshift prior based on the mean redshifts of the corresponding protoclusters (see Section~\ref{ssec:sfrd}).
For galaxies identified with a single emission line, we confirm that the SED fitting assuming the detected line is [O\textsc{iii}]$\lambda5007$ yields a better fit with smaller $\chi^2$ than assuming it is either H$\alpha$ or [O\textsc{ii}]$\lambda\lambda3726,3729$.
The parameters used in the SED fitting process are detailed in Table~\ref{tab:sedparam}.
As will be discussed in the results (Section~\ref{ssec:metal}), it should be noted that the metallicity derived from the emission line ratio falls within the metallicity range derived from the SED fitting.
Moreover, when the metallicity is fixed to that derived from the emission line ratio, other physical parameters such as stellar mass and SFR do not change significantly within the uncertainties.

\begin{table*}
    \centering
    \caption{SED fitting parameters}
    \begin{tabular}{ccc}
        \toprule
       Parameter & Prior & Range \\
        \midrule
       Attenuation $(A_V)$ / mag & Uniform & [0.0, 5.0]\\
       Time since star formation / Gyr & Uniform & [0.01, 14.0]\\
       Total formed stellar mass / M$_\odot$ & logarithmic & [$10^{5.0}$, $10^{13.0}$]\\
       Metallicity /  $Z_\odot$ & logarithmic & [0.01, 1.0]\\
       Timescale of star formation decreasing $(\tau)$ / Gyr & Uniform & [0.001, 10.0]\\
       Ionisation parameter / $\log{U}$ & Uniform & [-4, -2] \\
       \bottomrule
    \end{tabular}
    \label{tab:sedparam}
\end{table*}

We construct the catalogue by cross-matching the protocluster members identified by PCFNet in the HSC-SSP S20A data with the joint catalogue using sky coordinates. 
For galaxies with successful matches, we adopt the $uu^*grizyJHK$ photometry from the joint catalogue for the SED fitting. 
For unmatched sources, we instead use the $grizy$ photometry from the HSC-SSP S20A data. 
All photometry is based on PSF-convolved \texttt{forced} measurements with a $2.0''$ aperture calculated for a seeing of $0.59''$, consistent with our target selection procedure.
Note that there is no near-infrared observation in the DEEP2-3\_ID2 region, so all galaxies in the region are fitted with only $ugrizy$ bands, which may lead to larger uncertainties in the estimated age or dust attenuation.

The photometry of the $JHK$ bands in the joint catalogue is measured using the HSC pipeline \citep{Bosch2018}.
However, concerns had been raised that the error evaluation might be underestimated because it does not account for the covariance of background noise.
Therefore, we re-evaluate it as the approach established by \citet{Labbe2003, Gawiser2006}.
We first place 10,000 random points in the $JHK$-band images resampled to the same pixel scale as HSC, avoiding overlaps with detected objects.
Then, we calculate the flux of background noise within apertures of radii ranging from $2$ to $50$ pixels (corresponding to $0.34\text{--}8.4''$) around each random point.
From the flux distribution in each aperture, we calculate the median absolute deviation (MAD) for each image, assuming a Gaussian distribution, and use $1.48 \times \mathrm{MAD}$ as the $1\sigma$ uncertainty.
We then fit the following function to the aperture area for each image:
\begin{equation}
    \sigma_N = \alpha N^{\beta},
\end{equation}
where $\sigma_N$ is the $1\sigma$ uncertainty of the background noise within an aperture of $N$ pixels, and $\alpha$ and $\beta$ are fitting parameters.
Using the $\alpha$ and $\beta$ obtained for each band image, we calculate the $1\sigma$ uncertainty of the background noise within a $2.0''$ aperture, which is equivalent to the aperture size of photometry in the joint catalogue.
The average scaling ratio obtained for each band image is then used to rescale the uncertainties in the $JHK$ bands by a factor of 2.9.

\label{ssec:metalest}

\section{Results}
\label{sec:results}
\subsection{Protocluster identification}
\label{ssec:pcdetect}

\begin{figure*}
    \centering
    \includegraphics[width=0.45\linewidth]{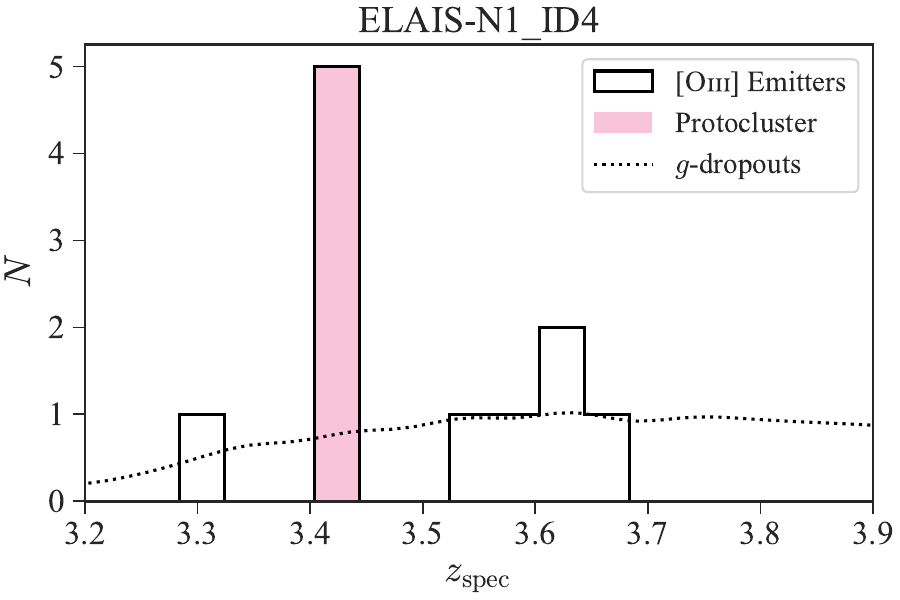}
    \includegraphics[width=0.45\linewidth]{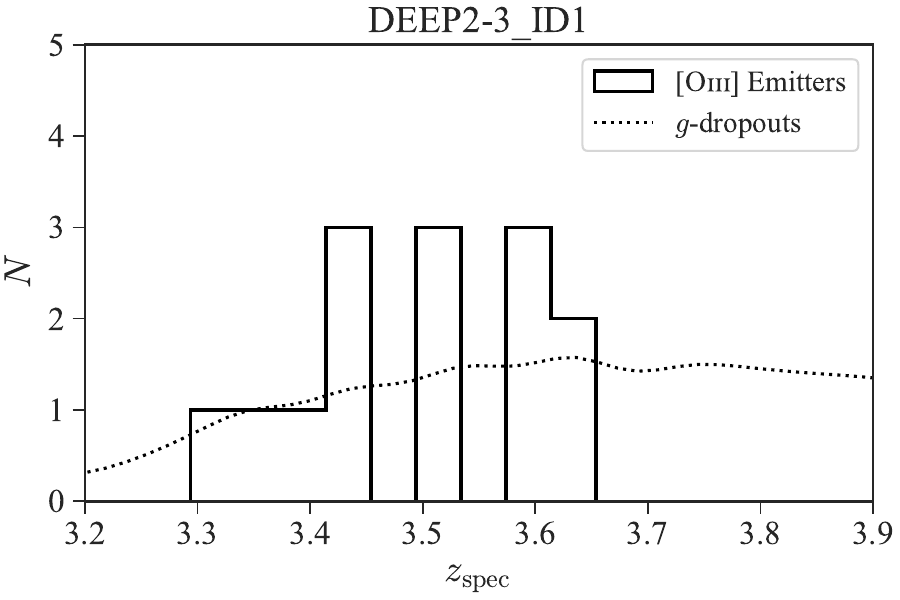}\\
    \includegraphics[width=0.45\linewidth]{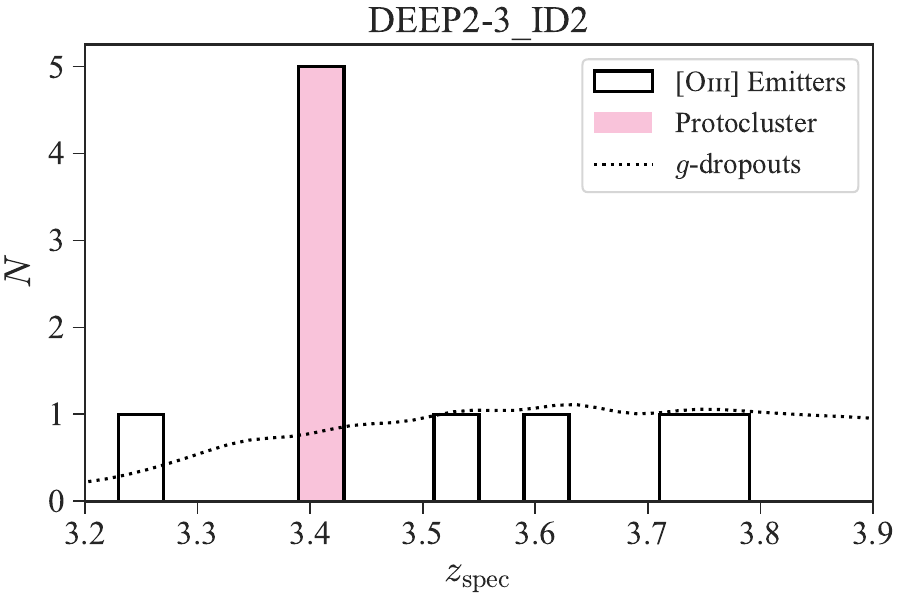}
    \includegraphics[width=0.45\linewidth]{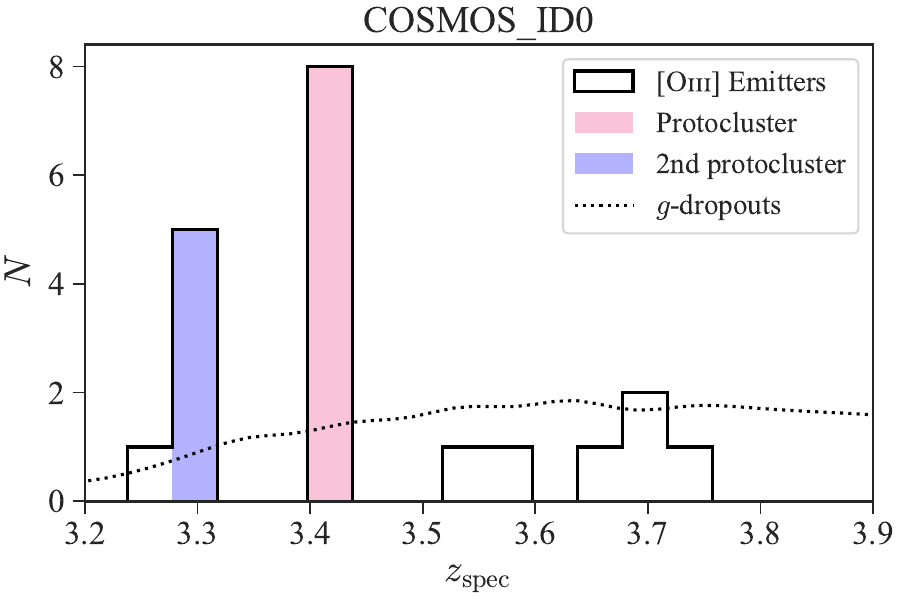}
    \caption{Redshift distribution of each protocluster candidate region. The black histogram represents the redshift distribution of spectroscopically identified [O\textsc{iii}] emitters with a window size of $\Delta z=0.04$. 
    The coloured shaded regions indicate the identified protoclusters.
    The dotted line represents the expected uniform distribution of $g$-dropout galaxies, normalised to the total number of objects in each region.}
    \label{fig:zdist}
\end{figure*}

Fig.~\ref{fig:zdist} presents the spectroscopic redshift distributions with redshift window $\varDelta z=0.04$ for each observed region.
Significant peaks, five times higher than the expected uniform distribution of $g$-dropout galaxies, are clearly observed in the ELAIS-N1\_ID4, DEEP2-3\_ID2, and COSMOS\_ID0 fields.
The COSMOS\_ID0 field exhibits two redshift peaks at $z=3.316$ and $z=3.425$.
Their separation, $\Delta z=0.109$, corresponds to a rest-frame velocity difference of $\Delta v \simeq c\Delta z/(1+z)\sim7.5\times10^3\,\mathrm{km\,s^{-1}}$ at $z\sim3.37$, far exceeding the typical internal velocity width of a single protocluster \citep[e.g.][]{Toshikawa2020}.
We therefore treat them as independent structures.
These peaks are considered to represent protoclusters on the basis of their high overdensities and are hereafter referred to as ELAIS-N1\_ID4\_PC1, DEEP2-3\_ID2\_PC2, COSMOS\_ID0\_PC3, and COSMOS\_ID0\_PC4.
The basic properties of the identified protoclusters are summarised in Table~\ref{tab:pc_basic_properties}.
The identified protoclusters have median redshifts of $z=3.408$ (ELAIS-N1\_ID4\_PC1), $z=3.403$ (DEEP2-3\_ID2\_PC2), $z=3.425$ (COSMOS\_ID0\_PC3), and $z=3.316$ (COSMOS\_ID0\_PC4).
In contrast, the DEEP2-3\_ID1 field shows a broad redshift distribution, with no more than three galaxies, corresponding to an overdensity of only about twice the uniform expectation, in any given bin; therefore, we conclude no protoclusters are detected in the field.
Finally, ELAIS-N1\_ID4\_PC1, DEEP2-3\_ID2\_PC2, COSMOS\_ID0\_PC3, and COSMOS\_ID0\_PC4 have five, five, eight, and five spectroscopically identified members, respectively.
A total of 23 protocluster member galaxies and 34 field galaxies are spectroscopically identified.

\begin{table*}
    \centering
    \caption{Basic properties of the identified protoclusters.}
    \begin{tabular}{ccccc}
        \toprule
        ID & R.A. (J2000) & Decl. (J2000) & $z_\mathrm{spec}^a$ & $N_\mathrm{mem}$\\
        \midrule
        ELAIS-N1\_ID4\_PC1 & 16:12:29.94 &  +55:59:28.79 & 3.408 &  5 \\
        DEEP2-3\_ID2\_PC2 & 23:29:54.44 &  +00:43:09.41 & 3.403 & 5 \\
        COSMOS\_ID0\_PC3 & 09:57:27.49 &  +00:58:47.53 & 3.425 & 8 \\
        COSMOS\_ID0\_PC4 & 09:57:33.59 &  +00:58:30.14 & 3.316 & 5 \\
        \bottomrule
    \end{tabular}
    \medskip
    \begin{minipage}{\linewidth}
    \textit{Notes.} $^a$ Median spectroscopic redshift of member galaxies. 
    \end{minipage}
    \label{tab:pc_basic_properties}
\end{table*}

Our spectroscopic observations provide a direct test of PCFNet's protocluster detection capability.
PCFNet achieves a purity of $69\pm4$\% for protocluster detection when a threshold of $\sigma_{\mathrm{th}}=2.5$ is applied to protocluster member galaxies \citep{Takeda2024}.
Our observed success rate of 75\% (3/4) is in agreement with this predicted purity, falling well within the expected range given the statistical uncertainties.
This consistency provides validation that PCFNet's performance on simulation data translates effectively to real observational data, confirming its reliability for identifying genuine protoclusters.
We also assess PCFNet's performance at the individual galaxy level by comparing its member predictions with our spectroscopic identifications.
Among the galaxies predicted by PCFNet to have high protocluster membership significance ($\sigma_\mathrm{prob}>2.5$), 15 out of 38 are actually protocluster members.
This is consistent with the precision predicted by PCFNet \citep[44\%;][]{Takeda2024} in the range of the confidence interval (32--48\%) of the binomial distribution \citep{Wilson1927}, indicating that PCFNet's predictions are statistically reliable.
The overall detection rate of protocluster members among the [O\textsc{iii}]-detected galaxies is 40\% (23/57), which is notably higher than the rates achieved by previous LBG-based detection methods \citep[typically $\sim10$--25\%, e.g.,][]{Toshikawa2016,Toshikawa2020, Toshikawa2025}.

\subsection{Properties of protocluster members and others}
\label{ssec:sedfit_properties} 
\begin{figure}
    \centering
    \includegraphics[width=\linewidth]{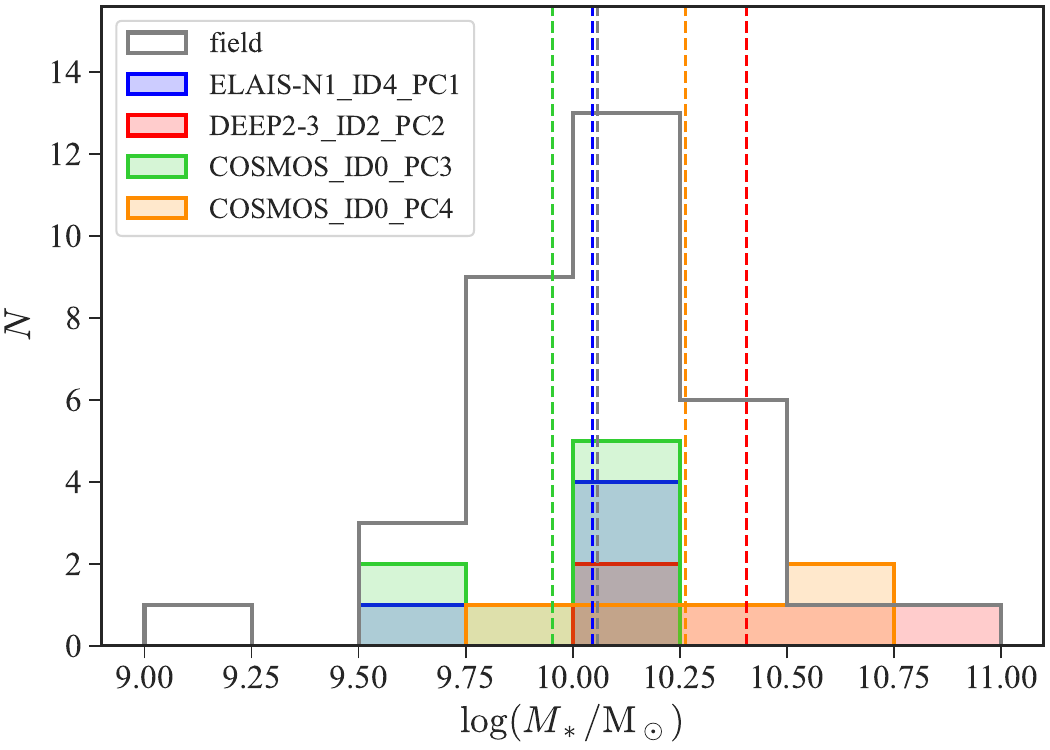}
    \caption{Histogram of the stellar mass for protocluster members (blue: ELAIS-N1\_ID4\_PC1, red: DEEP2-3\_ID2\_PC2, light green: COSMOS\_ID0\_PC3, orange: COSMOS\_ID0\_PC4) and field galaxies (i.e. [O\textsc{iii}] detected non-members, grey).
    The vertical dashed lines indicate the mean stellar mass of each protocluster member and the field galaxies.
    }
    \label{fig:stellarmass}
\end{figure}

We apply SED fitting for spectroscopically identified objects (see Section~\ref{ssec:sed_fitting} for details) to derive their physical properties.
Fig.~\ref{fig:stellarmass} shows the stellar mass distribution of protocluster members and field galaxies.
The members of DEEP2-3\_ID2\_PC2 and COSMOS\_ID0\_PC4 may exhibit a bias towards higher stellar masses. 
We statistically assess whether the stellar masses of DEEP2-3\_ID2\_PC2 and COSMOS\_ID0\_PC4 are biased towards the high-mass compared to field galaxies by performing Welch's t-test \citep{Welch1947}.
Note that DEEP2-3\_ID2 lacks $JHK$-band data, which may lead to larger uncertainties in the stellar mass estimates.
To address this, we compare the members of COSMOS\_ID0\_PC4 with field samples using the results of SED fitting that includes $JHK$-band data, while for DEEP2-3\_ID2\_PC2, we compare the SED fitting results based solely on $grizy$ bands between the members and the field sample.
The $p$-values of the null hypothesis that the stellar mass distributions of DEEP2-3\_ID2\_PC2 and COSMOS\_ID0\_PC4 are less than or equal to that of field galaxies are $0.29^{+0.28}_{-0.20}$ and $0.15^{+0.20}_{-0.11}$, respectively.
The uncertainties are estimated from 500 Monte Carlo realisations of the stellar mass from the posterior distribution of the SED fitting.
This indicates that neither protocluster shows a statistically significant ($p<0.05$) bias towards higher stellar masses compared to field galaxies.
We conclude that the stellar mass distributions in the four protoclusters are not significantly different from that in the field.
However, it should be noted that the small number of detected protocluster members ($N = 5$) imposes limitations on the statistical power of the tests, making it difficult to detect any potential bias towards higher stellar masses compared to field galaxies.

Previous studies \citep[e.g.][]{Galbiati2024} have shown that protoclusters tend to host more massive galaxies, and \citet{SunH2024} reported a top-heavy stellar mass function for star-forming galaxies in a protocluster at $z=2.5$.
These systems are suggested to be relatively evolved environments where galaxy growth is actively progressing, possibly fuelled by continuous gas accretion or frequent mergers. 
In contrast, our sample does not show such a pronounced bias towards massive galaxies. 
This could indicate that these are either intrinsically lower-mass protoclusters (see Sec.~\ref{ssec:halomass}) or are observed at an earlier stage of their assembly, where significant mass build-up has not yet occurred. 
It is also possible that massive galaxies in our protoclusters are dusty and missing in our sample due to the LBG selection, but this cannot be conclusively determined from the current results alone.

\begin{figure}
    \centering
    \includegraphics[width=\linewidth]{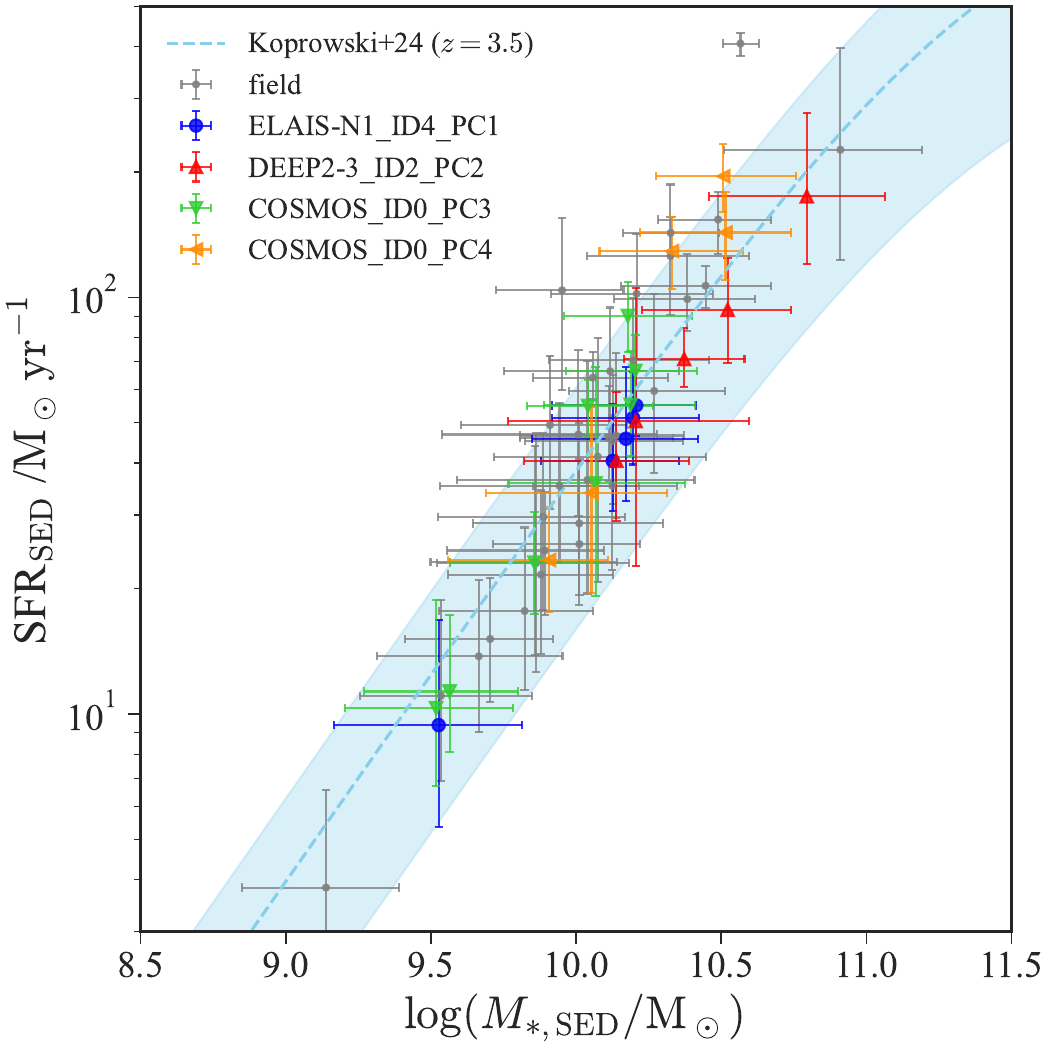}
    \caption{Scatter plots of stellar mass vs. SFR.
    The protocluster members of ELAIS-N1\_ID4\_PC1 (blue filled circles), DEEP2-3\_ID2\_PC2 (red triangles), COSMOS\_ID0\_PC3 (light green inverted triangles), and COSMOS\_ID0\_PC4 (orange right triangles) are indicated with error bars that represent a 68\% range estimated by SED fitting.
    The field galaxies (grey points, i.e. [O\textsc{iii}]-detected non-members) are shown in the same manner. The blue dashed line indicates the main sequence of SFG at $z=3.5$ \citep{Koprowski2024}, and the shaded region shows the uncertainty.
}
    \label{fig:ms}
\end{figure}

Fig.~\ref{fig:ms} presents the relationships between stellar mass and SFR.
Protocluster members are generally located on the star-forming main sequence at $z\sim3.5$, consistent with \citet{Koprowski2024}. 
No starburst galaxies are identified in our protocluster members.
Some previous studies \citep[e.g.][]{Galbiati2024} indicate that protocluster members lie on the main sequence, which is consistent with our results.
Other studies \citep[e.g.][]{Li2024} suggest that protocluster members might have slightly elevated SFRs compared to the main sequence. However, in our protocluster sample, no significant difference is observed, although this may be due to uncertainties inherent in the SED fitting process.

\subsection{Halo mass of protoclusters}
\label{ssec:halomass}
Estimating the masses of the most massive halos associated with the identified protoclusters at the observed epoch is useful for inferring their subsequent evolution into clusters. 
In previous studies, various methods have been employed to estimate protocluster halo masses \citep[e.g.][]{Overzier2009,Lemaux2014,Wang2016}, often probing different physical scales. 
In this work, we focus on the mass of the most massive halo associated with each protocluster, corresponding to the core galaxy that traces the main branch of the merger tree. 
This main-halo mass is distinct from the total or large-scale protocluster mass measured over extended regions in many observational studies. 
We estimate this quantity using the stellar-to-halo mass ratio (SHMR) derived from PCcone, which was used to train PCFNet. 
This definition is consistent with that commonly adopted in simulations for tracing the halo-mass growth of protoclusters \citep[e.g.][]{Chiang2013,Montenegro-Taborda2023,Lim2024}.
The core galaxy possesses the most massive halo mass and stellar mass within the protocluster, and its halo mass can be inferred from the stellar mass through the SHMR of the PCcone.
To derive the SHMR, we first select galaxy groups in PCcone that are predicted to be protoclusters by PCFNet.
It should be noted that these groups are not guaranteed to be actual protoclusters.
We fit a cubic function to the stellar mass and halo mass of the galaxy with the largest stellar mass of each group, deriving the relationship between stellar mass and halo mass.
We confirm that derived SHMR for the core galaxies does not differ significantly from the observed relation (see Appendix~\ref{sec:halomassestimation} for details).
We then estimate the halo mass by applying the ratio to the stellar mass of the galaxy with the largest stellar mass among the spectroscopically identified protocluster members.
It should be noted that this method has significant uncertainty when the core galaxy is either quiescent or dust-obscured and has not been spectroscopically identified \citep{Sillassen2024}.
The halo mass of ELAIS-N1\_ID4\_PC1, DEEP2-3\_ID2\_PC2, COSMOS\_ID0\_PC3, and COSMOS\_ID0\_PC4 is estimated to be $\log{M_\mathrm{halo}/\mathrm{M_\odot}}=12.4^{+0.38}_{-0.29}, 12.8^{+0.49}_{-0.31}, 12.4^{+0.39}_{-0.26}$ and $12.6^{+0.41}_{-0.28}$, respectively (see Table~\ref{tab:pc_properties}).
The uncertainties are estimated by propagating those in the stellar mass estimated from the SED fitting and in the SHMR fitting errors.
For reference, Table~\ref{tab:pc_properties} also lists the estimates of the total halo mass by summing the halo masses derived from the stellar masses of all spectroscopically identified members using the SHMR from $g$-dropouts in PCcone.

\begin{figure}
    \centering
    \includegraphics[width=\linewidth]{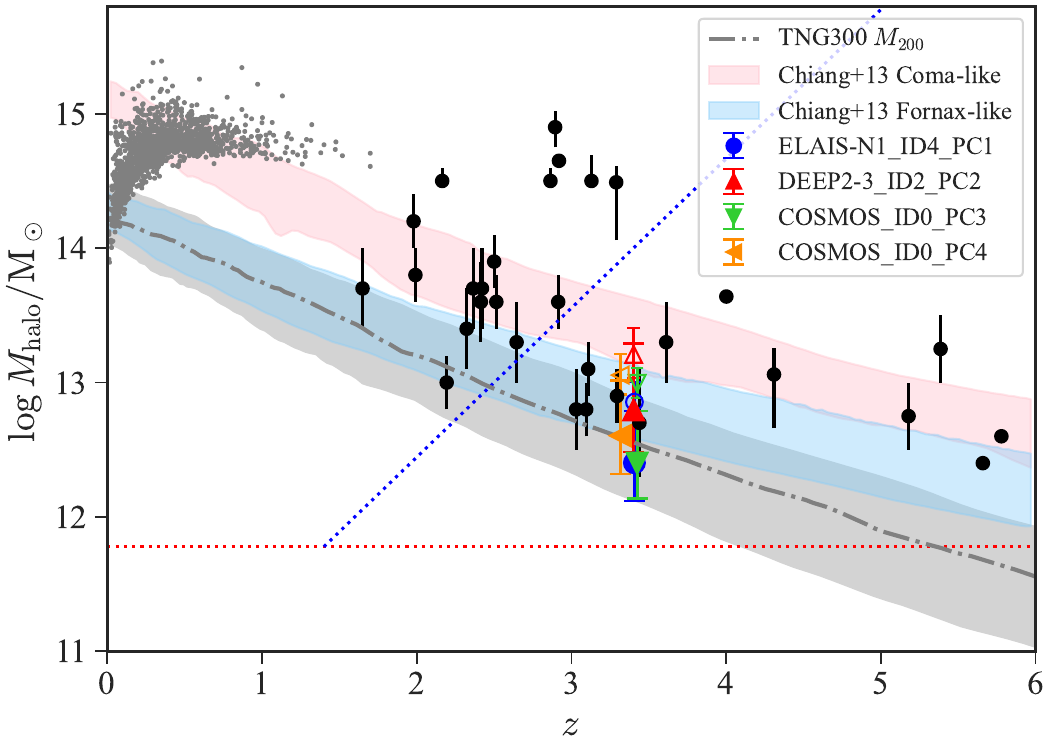}
    \caption{ The growth history of protocluster as revealed by their halo mass evolution as a function of redshift.
    Coloured filled symbols represent the main-halo masses of our protoclusters estimated using the SHMR derived from PCcone (symbols as in Fig.~\ref{fig:ms}).
    Open symbols indicate the total halo mass estimated by summing the halo masses of all spectroscopically identified members.
    Red and blue shaded regions show the halo mass evolution of Coma-like ($M_\mathrm{halo}^{z=0}>10^{15} \mathrm{M_\odot}$) and Fornax-like ($M_\mathrm{halo}^{z=0}=1.37\text{-}3\times10^{14} \mathrm{M_\odot}$) protoclusters from \citet{Chiang2013}.
    Grey dash-dotted line and shaded region indicate the halo mass evolution and uncertainty of $M_{200}>10^{14}\mathrm{M_\odot}$ protoclusters in TNG300 simulation \citep{Montenegro-Taborda2023}.
    The blue and red dotted lines indicate the boundaries of gas cooling regimes proposed by \citet{Dekel2006}.       
    The black circles indicate the protoclusters in the literature \citep{Sillassen2024,Jin2024,Miller2018,Oteo2018,Morishita2025,Sun2024,Helton2024,Lemaux2014,Cucciati2014,Venemans2007,Daddi2022}.
    The grey circles show nearby clusters in the literature \citep{Planck2016,Bocquet2019}.
    Note that these are based on heterogeneous definitions and estimations of mass and are not necessarily directly comparable to the main-halo mass estimates in this study. 
    }
    \label{fig:halo_mass_evolution}
\end{figure}

Fig.~\ref{fig:halo_mass_evolution} shows the redshift evolution of the most massive halo associated with protoclusters. 
The literature values included for comparison are based on heterogeneous definitions and estimations of mass, including the most massive halo mass, the sum of individual halo masses, and masses inferred over larger spatial scales. 
No aperture correction is applied to these literature measurements. 
The figure also shows the mass evolution of the most massive single halo within protoclusters in the TNG300 simulation and the model by \citet{Chiang2013} for different halo masses at $z=0$.
Based on simulated mass accretion histories \citep{Montenegro-Taborda2023}, the halo masses of our protocluster sample fall within the range expected for average protoclusters at this redshift.
This result demonstrates that PCFNet can identify typical-mass protoclusters, avoiding a bias towards only massive systems.
The result is also confirmed even if using other SHMR estimation \citep[e.g.,][see Appendix~\ref{sec:halomassestimation}]{Shuntov2022}.
Moreover, all of the identified protoclusters are likely in the cold-in-hot regime \citep{Dekel2006} on the halo mass-redshift plane, allowing the presence of cold gas inflows penetrating hot gas halos.

\begin{table*}
    \centering
    \caption{Derived physical properties of the identified protoclusters.}
    \begin{tabular}{cccccc}
        \toprule
        ID & $\log{M_{*,\mathrm{max}}/\mathrm{M_\odot}}^a$& $\log{M_\mathrm{halo}/\mathrm{M_\odot}}^b$ & $\log{M_\mathrm{halo,tot}/\mathrm{M_\odot}}^c$ & $V/10^2\mathrm{cMpc^3}$& SFRD /$\mathrm{M_\odot\,yr^{-1}\,cMpc^{-3}}$\\
        \midrule
        ELAIS-N1\_ID4\_PC1 & $10.21^{+0.21}_{-0.29}$& $12.4^{+0.4}_{-0.3}$& $12.9^{+0.1}_{-0.1}$ &$6.6^{+1.3}_{-1.2}$&$2.0^{+1.0}_{-0.6}$\\
        DEEP2-3\_ID2\_PC2 & $10.79^{+0.27}_{-0.34}$ & $12.8^{+0.5}_{-0.3}$& $13.2^{+0.2}_{-0.2}$&$8.2^{+1.4}_{-1.3}$&$2.9^{+1.4}_{-0.9}$\\
        COSMOS\_ID0\_PC3 & $10.20^{+0.21}_{-0.24}$ & $12.4^{+0.4}_{-0.3}$&$13.0^{+0.1}_{-0.1}$& $11.0 ^{+1.5}_{-1.5}$&$2.1^{+0.9}_{-0.6}$\\
        COSMOS\_ID0\_PC4 & $10.51^{+0.23}_{-0.29}$ & $12.6^{+0.4}_{-0.3}$&$13.1^{+0.2}_{-0.1}$& $18.3 ^{+1.8}_{-1.8}$&$1.6^{+0.6}_{-0.4}$\\
        \bottomrule
    \end{tabular}
    \medskip
    \begin{minipage}{\linewidth}
    \textit{Notes.} $^{a}$ Stellar mass of the member, which is the most massive in the protocluster. $^{b}$ Halo mass estimated using the SHMR computed from PCcone. $^{c}$ Total halo mass estimated by summing the halo masses of all spectroscopically identified members.
    \end{minipage}
    \label{tab:pc_properties}
\end{table*}

\subsection{SFRD of protoclusters}
\label{ssec:sfrd}
\begin{figure*}
    \centering
    \includegraphics[width=0.45\linewidth]{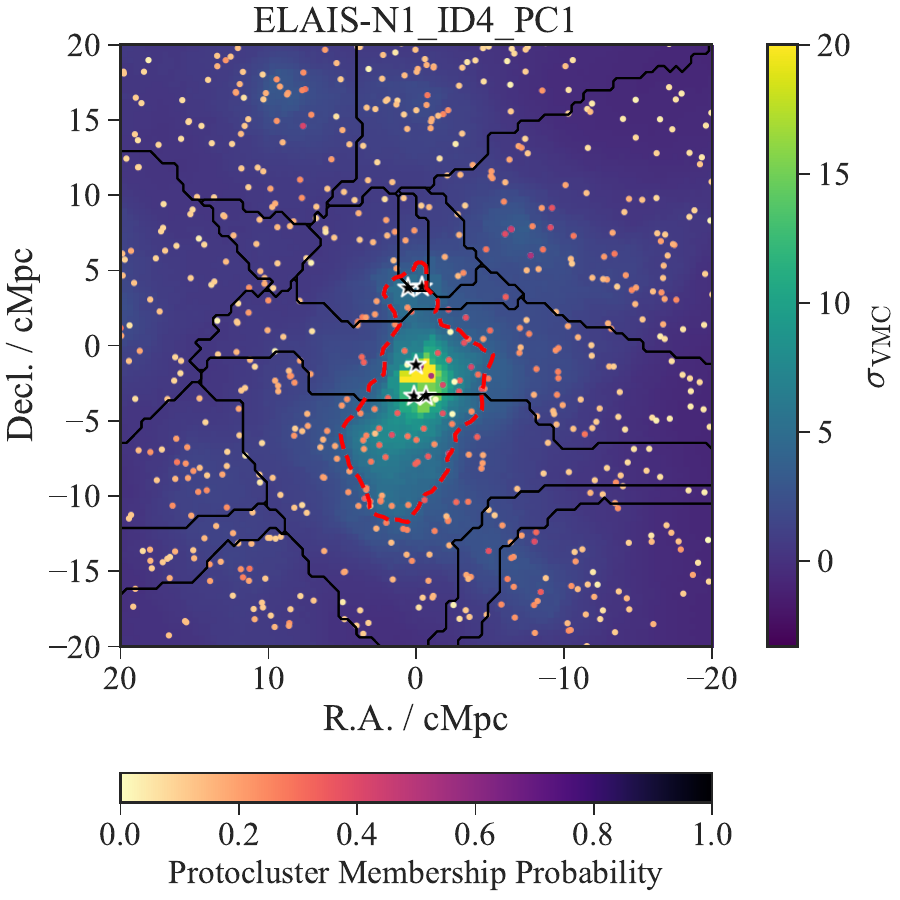}
    \includegraphics[width=0.45\linewidth]{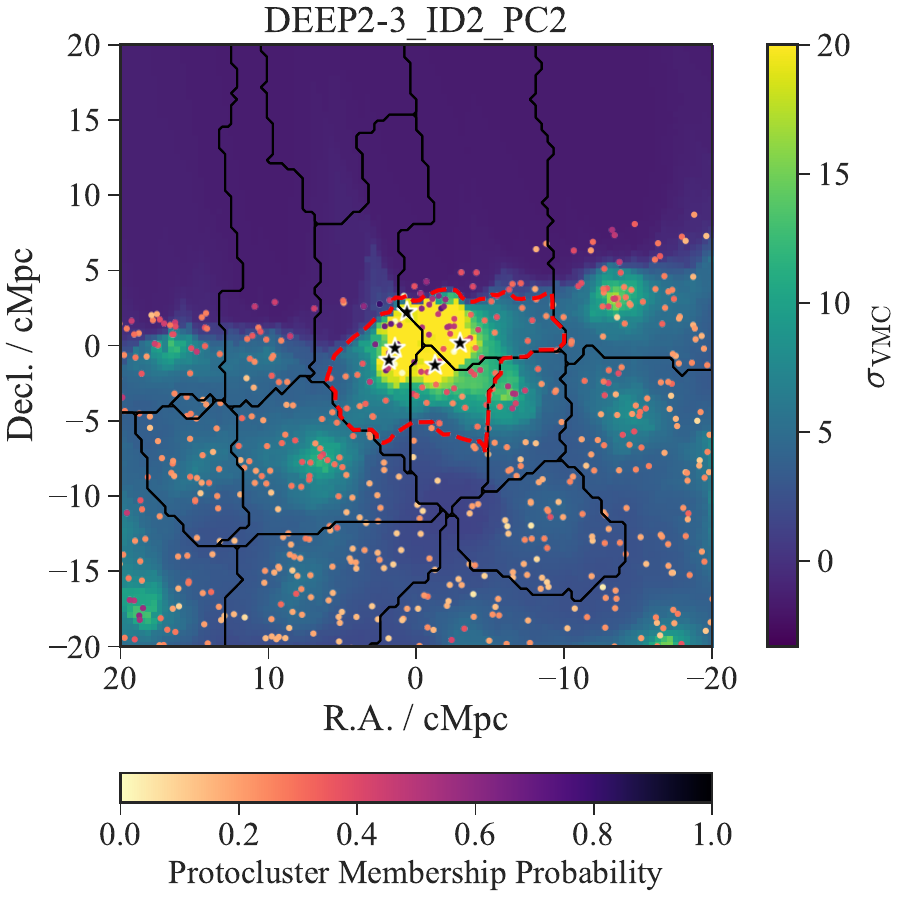}\\
    \includegraphics[width=0.45\linewidth]{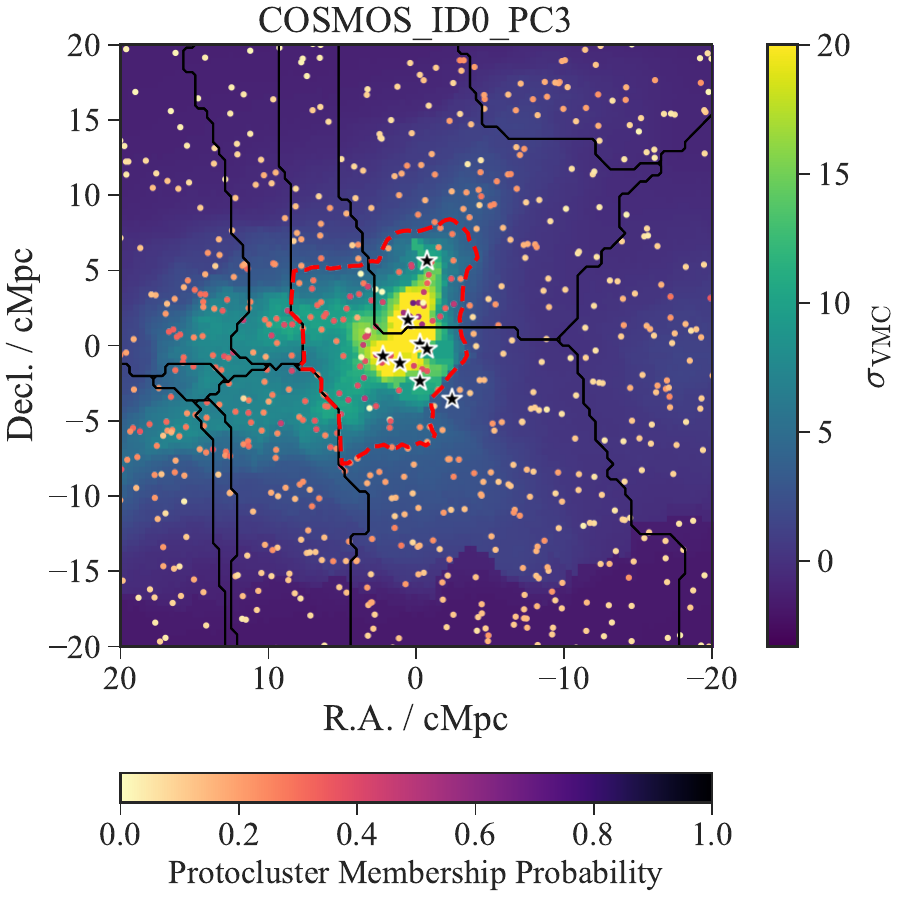}
    \includegraphics[width=0.45\linewidth]{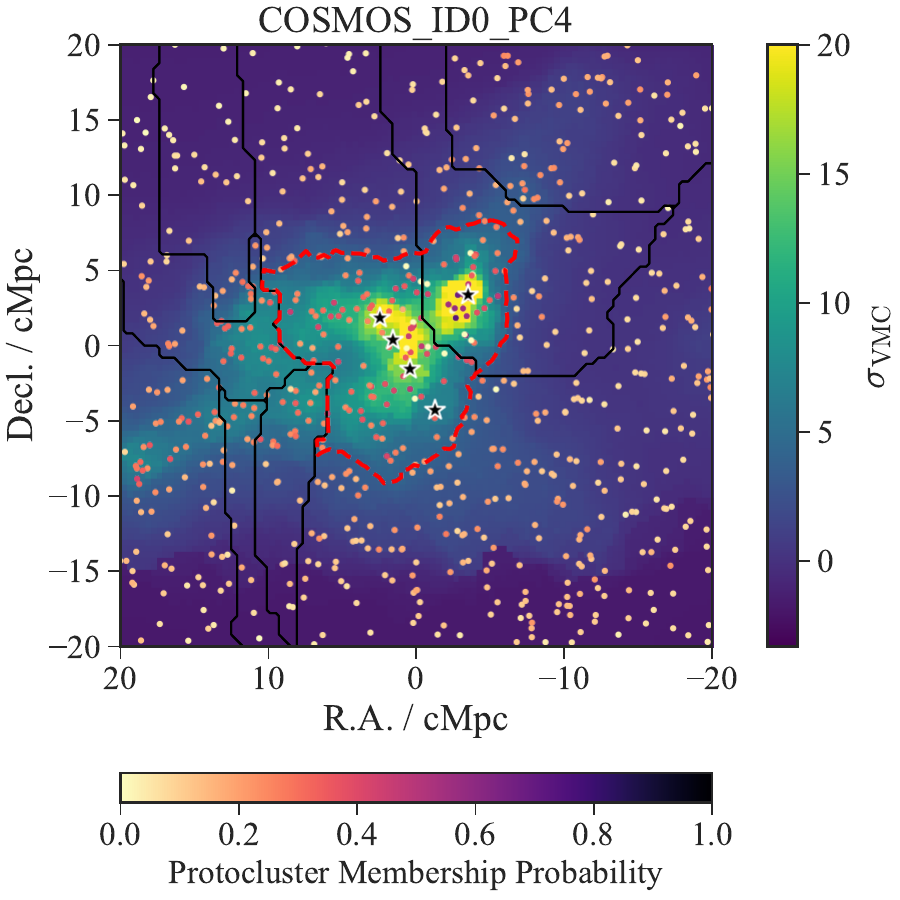}
    \caption{Sky distribution of $g$-dropout galaxies with protocluster member probability predicted by PCFNet for ELAIS-N1\_ID4\_PC1 (top left), DEEP2-3\_ID2\_PC2 (top right), COSMOS\_ID0\_PC3 (bottom left), and COSMOS\_ID0\_PC4 (bottom right).
    Coloured circles and black stars represent the $g$-dropout galaxies and spectroscopically identified protocluster members, with their colours indicating the probability of being a protocluster member as predicted by PCFNet.
    Background-filled contours show the mean density map $\bm{\sigma}_\mathrm{VMC}$ derived by the VMC approach.
    Black lines indicate the watershed segmentation boundaries, while red dashed lines indicate the estimated protocluster regions defined from the density threshold and watershed segmentation.}
    \label{fig:density_map}
\end{figure*}

In principle, an accurate measurement of the SFRD of a protocluster requires a nearly complete spectroscopic census of its members.
In practice, however, such completeness is difficult to achieve.
We therefore use the PCFNet membership probabilities to construct a probabilistic estimate of the total SFR and SFRD.
To determine the protocluster region, we employ a Voronoi Tessellation Monte Carlo (VMC) approach \citep{Lemaux2017}. 
While previous studies \citep[e.g.][]{Lemaux2017, Hung2020, Staab2024} have typically drawn realisations from photo-$z$ probability distribution functions, our method instead samples candidate member galaxies according to their membership probabilities, $p_i$, predicted by PCFNet.
Here, $p_i$ denotes the predicted probability that the $i$-th galaxy belongs to a protocluster somewhere along the line of sight, rather than specifically to the target protoclusters. 
In the present analysis, we assume that the target protoclusters are the dominant overdense structure along the line of sight and therefore use $p_i$ as a proxy for the membership probability of the target protoclusters. 
In each realisation, we generate a uniform random number $u_i \sim \mathcal{U}(0,1)$ and select the galaxy as a candidate member if $u_i < p_i$.
For spectroscopically confirmed members, the probability is fixed at 1, while for spectroscopically identified non-members, it is set to zero.
This spectroscopic assignment is made separately for each target protocluster, allowing the VMC density maps to differ even for structures along the same line of sight.
We then perform a two-dimensional Voronoi tessellation for selected candidate members on the sky-projected plane at the redshift of the protocluster and compute the cell corresponding to each galaxy. 
The local density is defined as the inverse of the cell area.
This procedure is repeated 100 times, and the results are averaged on a $0.4\,\mathrm{cMpc}$ grid to construct a density map $\bm{\rho}_\mathrm{VMC}$ over a $40 \times 40\,\mathrm{cMpc^2}$ region centred on the protocluster.
To estimate the field density and its variance, we randomly sample 1000 points from the regions of the HSC-SSP DEEP layer for which PCFNet provides protocluster member probabilities, and derive the field density distribution, $\bm{\rho}_\mathrm{field}$. 
We define protocluster regions as those with densities exceeding a threshold given by
\begin{gather}
\rho_\mathrm{th}=\mathrm{Med}(\bm{\rho}_\mathrm{field})+4\mathrm{MAD}(\bm{\rho}_\mathrm{field}),
\end{gather}
where $\mathrm{Med}(\bm{\rho}_\mathrm{field})$ and $\mathrm{MAD}(\bm{\rho}_\mathrm{field})=\mathrm{Med}(|\bm{\rho}_\mathrm{field}-\mathrm{Med}(\bm{\rho}_\mathrm{field})|)$ denote the median and the median absolute deviation of the field density distribution, respectively.
As the PCFNet-based selection may undesirably include galaxies associated with protoclusters at different redshifts, it is necessary to exclude such cases from our analysis.
We achieve this by applying watershed segmentation \citep{Vincent1991} to the density map. 
In watershed segmentation, local maxima in the density map are identified as markers, except for peaks with densities of $\leq 0.01\,\mathrm{cMpc}^{-2}$.
The sign-flipped density map is then flooded from each marker, and the regions are divided at the watershed lines where the floods from different markers meet.
These calculations are performed using the \texttt{scikit-image} library \citep{skimage2014}.
We then discard any segments that lack spectroscopically confirmed members from the protocluster region. 
Although this approach may overlook substructures within the protoclusters, we adopt it to ensure a more conservative analysis.
It should be noted that this assumption may break down if other protoclusters are aligned along the same line of sight. 
In such cases, this method may not effectively exclude those additional structures, potentially leading to overestimated volumes and total SFRs.
Fig.~\ref{fig:density_map} shows the VMC density maps where the significance is defined as $\bm{\sigma}_\mathrm{VMC}=(\bm{\rho}_\mathrm{VMC} - \mathrm{Med}(\bm{\rho}_\mathrm{field}))/\mathrm{MAD}(\bm{\rho}_\mathrm{field})$, and red contours mark the defined protocluster regions.
One spectroscopically confirmed member of COSMOS\_ID0\_PC3 lies outside the watershed-defined protocluster region.
This does not imply an inconsistency: spectroscopic membership is defined from the redshift peak,
whereas the protocluster region is defined independently from the projected density map.
The object is likely associated with a lower-density outer component or substructure that is not enclosed by our conservative watershed boundary.
It is important to note that DEEP2-3\_ID2\_PC2 is located at the edge of the survey region \citep[see Section 5.2 of][]{Takeda2024}, resulting in the absence of detected galaxies in the northern part of this protocluster, which might give large uncertainties in the estimates of its volume.
Additionally, COSMOS\_ID0\_PC3 and COSMOS\_ID0\_PC4 are closely aligned along the line of sight, so the assumption that the target PC is the dominant structure may be less reliable for these systems. 
Their candidate protocluster members can therefore overlap, which may result in overestimated volumes for these structures.

For the galaxies within a protocluster region determined in this manner, we derive the total SFR of the protoclusters by summing the SFRs of Monte Carlo-selected protocluster members, where each galaxy is selected according to its PCFNet membership probability.
For galaxies without spectroscopic redshifts, we estimate their SFRs by the SED fitting with a Gaussian prior centred on the redshift of the corresponding protocluster, $z_\mathrm{pc}$, with a width of $\sigma_{z}=0.2$. 
We also fix the parameters to typical values for LBGs, $A_V=1.0$, $\log U=-2.0$, and $Z=0.2Z_\odot$ (e.g., \citealt{Grazian2017,Alvarez-Marquez2019}).
Because this procedure explicitly assumes that the galaxy lies at $z_\mathrm{pc}$, forcing the fit for galaxies whose photometry is clearly inconsistent with the protocluster redshift can lead to catastrophic and unphysical SFR estimates. 
We therefore apply the criteria $\chi^2_{\nu,\mathrm{phot}}<3$ and $z_\mathrm{phot,16}-0.02 < z_\mathrm{pc} < z_\mathrm{phot,84}+0.02$ only as a quality-control step for the SFR estimation at fixed $z_\mathrm{pc}$.
Here, $\chi^2_{\nu,\mathrm{phot}}$ is the reduced chi-square of the best-fit SED model derived by \textsc{Bagpipes}, and $z_\mathrm{phot,16}$ and $z_\mathrm{phot,84}$ are the 16th and 84th percentiles of the photometric-redshift probability distribution function, respectively. 
We perform 1000 Monte Carlo realisations of protocluster membership sampling and measure the mean and standard deviation of the total SFR across these realisations.

To calculate the SFRD, we estimate the volume of each protocluster.
Although protoclusters generally exhibit asymmetric shapes, for simplicity, we approximate the volume as $V=\tfrac{4}{3}\pi R_{xy}^2R_z$, where $R_{xy}=\sqrt{A/\pi}$ is the pseudo-radius derived from the area $A$ of the protocluster region defined above and $R_z$ is the line-of-sight radius.
Since it is difficult to accurately determine $R_z$ due to the small number of spectroscopically confirmed members, we assume spherical symmetry and set $R_z=R_{xy}$ for simplicity.
Note that this provides an order-of-magnitude estimate of the volume and should be interpreted with caution.
We then compute the total SFR and divide it by the estimated volume to obtain the SFRD.

We apply corrections for the contribution of low-luminosity galaxies and selection completeness to the derived SFRDs.
Following \citet{Staab2024}, we estimate the contribution of low-luminosity galaxies below our detection limit down to $M_\mathrm{UV}=-17$.
The correction factor is calculated as the ratio of the integrated luminosity function as a Schechter function from $M_\mathrm{UV}=-17$ to infinity to that from our detection limit ($M_\mathrm{UV}=-19.8$ corresponding to $i=26$ at $z\sim3.4$) to infinity:
\begin{gather}
    \scalebox{0.9}{\ensuremath{
    \mathrm{CF}_\mathrm{low\_lum} = \dfrac{
    \displaystyle \int_{-\infty}^{-17} \left[10^{0.4(M^\ast_\mathrm{UV} - M)}\right]^{\alpha+1}
    \, \exp\left[-10^{0.4(M^\ast_\mathrm{UV} - M)}\right] dM}{
    \displaystyle \int_{-\infty}^{-19.8}\left[10^{0.4(M^\ast_\mathrm{UV} - M)}\right]^{\alpha+1}
    \, \exp\left[-10^{0.4(M^\ast_\mathrm{UV} - M)}\right]
    dM},}}
\end{gather}
where $M^\ast_\mathrm{UV}$ and $\alpha$ are the characteristic magnitude and the faint-end slope of the Schechter function, respectively.
We adopt $M^\ast_\mathrm{UV}=-20.61^{+0.12}_{-0.14}$ and $\alpha=-0.16^{+0.25}_{-0.25}$ from the Schechter luminosity function of the $g$-dropout protocluster members \citep{Ito2020}, resulting in a correction factor of $0.24^{+0.06}_{-0.05}$ dex.
Note that this correction factor is lower than the 0.96 dex when using the field Schechter function \citep{Bouwens2015} employed by \citet{Staab2024}.
The selection completeness, which takes into account the galaxies that are missed during colour selection, of $g$-dropout galaxies is estimated to be 0.55 based on \citet{Harikane2022}, and we apply this correction as well.
The detection completeness at our detection limit ($i=26$) is approximately 1.0 according to \citet{Desprez2023}, so we do not apply any correction for this.

\begin{figure}
    \centering
    \includegraphics[width=\linewidth]{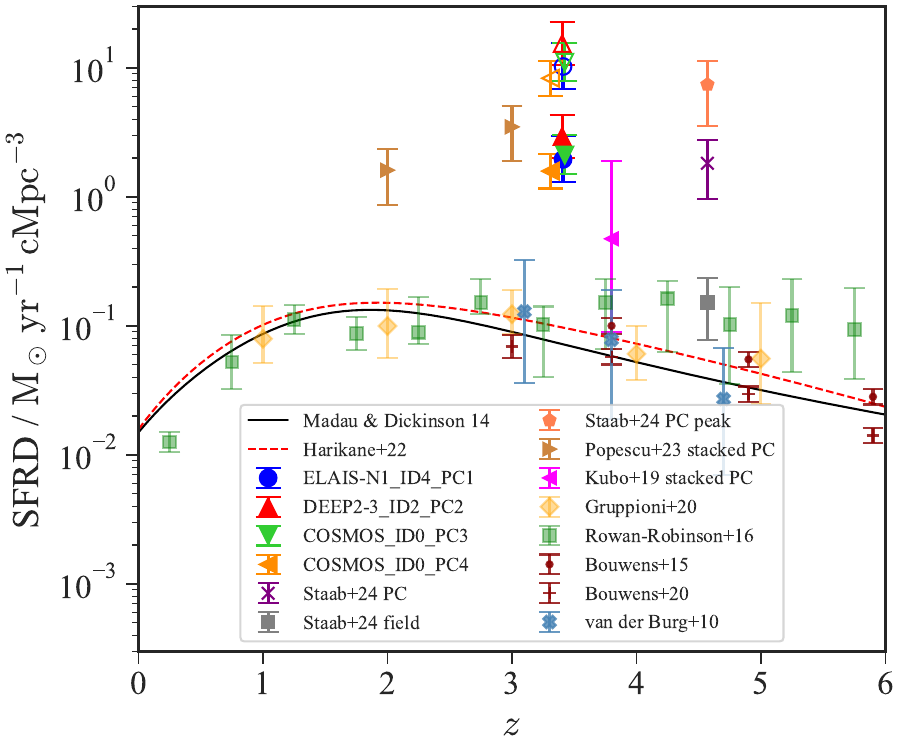}
    \caption{SFRD as a function of redshift. 
    Coloured filled and open symbols represent the SFRD of protoclusters estimated using the VMC approach with correction factors of 0.24 and 0.96 dex inferred from the luminosity functions of \citet{Ito2020} and \citet{Bouwens2015}, respectively (symbols as in Fig.~\ref{fig:ms}).
    Measurements on field data \citep{Gruppioni2020,Rowan2016,Bouwens2015,Bouwens2020,vanderBurg2010} are plotted as yellow diamonds, forest green squares, wine-red circles and pluses, and light blue crosses.
    The SFRD of the protocluster regions, protocluster peak regions, and field in \citet{Staab2024} are shown as a magenta cross, an orange pentagon, and a grey square, respectively.
    The orange triangles indicate \citet{Popescu2023} stacked protoclusters at $z=2.0$ and $3.0$ with a radius of $10.5\,\mathrm{cMpc}$.
    The pink triangle indicates the SFRD of the protocluster at $z=3.8$ by \citet{Kubo2019} assuming a radius of $10.2\,\mathrm{cMpc}$.
    The solid blue line and dashed red line represent the cosmic SFRD from \citet{Madau2014,Harikane2022}. 
    }
    \label{fig:sfrd}
\end{figure}

The $R_{xy}$ for ELAIS-N1\_ID4\_PC1, DEEP2-3\_ID2\_PC2, COSMOS\_ID0\_PC3 and COSMOS\_ID0\_PC4 are estimated to be $5.4^{+0.3}_{-0.4},\ 5.8^{+0.3}_{-0.3},\ 6.4^{+0.3}_{-0.3}$ and $ 7.6^{+0.2}_{-0.2}$ cMpc, respectively.
These $R_{xy}$ are notably smaller than the average Lagrangian radius $R_L \sim 9$ cMpc reported for protoclusters at $z\sim3.5$ by \citet{Chiang2017}.
The definition of $R$ used in this work differs from that in \citet{Chiang2017} and may reflect a region closer to the core of the protocluster.
The resulting SFRDs for ELAIS-N1\_ID4\_PC1, DEEP2-3\_ID2\_PC2, COSMOS\_ID0\_PC3 and COSMOS\_ID0\_PC4 are $2.0^{+1.0}_{-0.6},\ 2.9^{+1.4}_{-0.9},\ 2.1^{+0.9}_{-0.6}$ and $1.6^{+0.6}_{-0.4}\,\mathrm{M_{\odot}}\,\mathrm{yr}^{-1}\,\mathrm{Mpc}^{-3}$, respectively.
The uncertainties are estimated by propagating those in the total SFR and volume estimates.
Fig.~\ref{fig:sfrd} presents these SFRDs compared with previous studies.
Our measured SFRDs in the protocluster regions are significantly higher than the cosmic SFRD at similar redshifts based on UV luminosity functions \citep{Bouwens2015,Bouwens2020,vanderBurg2010}, IR measurements \citep{Gruppioni2020,Rowan2016}, and combined UV+IR compilations \citep{Madau2014}.
It should be noted that the methods to estimate the cosmic SFRD vary from study to study, the comparison should be interpreted with caution.
The SFRDs of our protoclusters are comparable to those reported for other protoclusters at $z\sim4.57$ by \citet{Staab2024} and at $z=3.8$ \citep{Kubo2019}, which is estimated to be $0.47^{+1.4}_{-0.38}\,\mathrm{M}_\odot\,\mathrm{yr}^{-1}\,\mathrm{Mpc}^{-3}$ when assuming a radius of 10.2 cMpc \citep[corresponding to a $5'$ aperture at $z=3.8$ with standard cosmology $(\Omega_m, \Omega_\Lambda, h) = (0.3, 0.7, 0.70)$;][]{Popescu2023}.
They are also consistent with those of stacked protoclusters at $z=2\text{--}3$ reported by \citet{Popescu2023}, assuming a radius of 10.5 cMpc.
These comparisons suggest that, despite the relatively small region and inferred lower halo mass than that in the literature, our protoclusters represent regions of intense star formation activity.
It should be noted that because the radius of our protocluster is small, we may be observing regions similar to the protocluster core, which could lead to elevated SFRD compared to the entire protocluster.
Note that the high SFRD is primarily driven by the high galaxy density rather than individual galaxies having elevated sSFR relative to the main sequence (see Section~\ref{ssec:sedfit_properties}).
However, it is generally difficult to accurately evaluate the volume occupied by protoclusters and the total star-formation activity within them, and for this reason, the estimation methods vary among the studies.

\subsection{Metallicity diagnostic from emission line ratio}
\label{ssec:metal}

We derive gas-phase metallicities for the spectroscopically confirmed galaxies using the flux ratio of [O\textsc{iii}]$\lambda$5007 to H$\beta$. 
These measurements enable us to characterise the chemical enrichment of both protocluster members and field galaxies.
To derive the average metallicity of each protocluster, we stack the spectra of member galaxies in each protocluster and field galaxies.
The target galaxies for this purpose are those in which the [O\textsc{iii}]$\lambda5007$ emission line is detected at $3\sigma$ or higher and the H$\beta$ line is expected to be detectable within the observed wavelength range, regardless of their detections or non-detections.
The stacked spectra are based on five, five, eight, and four member galaxies for ELAIS-N1\_ID4\_PC1, DEEP2-3\_ID2\_PC2, COSMOS\_ID0\_PC3, and COSMOS\_ID0\_PC4, respectively, and 31 field galaxies.
Each spectrum is normalised by the [O\textsc{iii}]$\lambda5007$ flux, and the stacked spectrum is computed at each wavelength using inverse-variance weighting based on the corresponding error spectrum.
The results do not change significantly, even when using simple average stacking without weighting.
To examine any mass-dependent differences in the field population, the field sample is further subdivided into low- and high-stellar-mass bins ($M_*<10^{10}\,\mathrm{M}_\odot$, $M_*\geq10^{10}\,\mathrm{M}_\odot$), and stacked separately.
The numbers of galaxies in the low- and high-mass bins are 12 and 19, respectively.
For the stacked spectra, we perform a simultaneous Gaussian profile fit to the [O\textsc{iii}] and H$\beta$ emission lines with fixed wavelength ratio using \texttt{astropy} \citep{astropy2022}.

The line ratio is then converted to metallicity using the [O\textsc{iii}]/H$\beta$ metallicity calibration presented by \citet{Sanders2025}.
It is important to note that this calibration exhibits low and high metallicity branches with a turnover at $12+\log{\mathrm{O/H}}\sim7.94$.
From the mass-metallicity relation of star-forming galaxies at $z\sim3.4$ \citep{Troncoso2014}, the stellar mass which corresponds to this turnover metallicity is approximately $M_*\sim10^{9.4}\,\mathrm{M}_\odot$.
Since all but one of the galaxies in our sample exceed this stellar mass, they can be considered to belong to the high-metallicity branch.
Therefore, when analysing the stacked spectra, we exclude galaxies with $M_*<10^{9.4}\,\mathrm{M}_\odot$ and estimate the metallicity from the high-metallicity branch.
Even considering the uncertainty in the calibration from \citet{Sanders2025} ($\sigma_{\mathrm{O/H}}=0.14$), we also performed the analysis using a conservative stellar mass threshold of $M_*=10^{9.9}\,\mathrm{M}_\odot$ ($12+\log{\mathrm{O/H}}=8.08$), but the results did not change significantly.
Additionally, it should be noted that the [O\textsc{iii}]$\lambda5007$/H$\beta$ ratio is sensitive to the ionisation parameter \citep[e.g.][]{Kewley2019}, which may differ between protocluster members and field galaxies \citep{Shimakawa2015}.
Galaxies with higher specific SFRs tend to emit more ionising photons, leading to higher ionisation parameters and potentially underestimating metallicity.
As discussed in Section~\ref{ssec:sedfit_properties}, both protocluster members and field galaxies are distributed along the star-forming main sequence, and no significant differences in sSFR are observed, suggesting that the differences in ionisation parameters are not substantial.
Moreover, we investigate how much [O\textsc{iii}]$\lambda$5007/H$\beta$ changes when varying the ionisation parameter while keeping the metallicity fixed using \textsc{cloudy} \citep{Gunasekera2025}.
We employ the BPASS v2.2.1 stellar population synthesis models \citep{Eldridge2017, Stanway2018}, adopting a Chabrier initial mass function \citep{Chabrier2003} with an upper mass cutoff of 300 $M_{\odot}$ and including the effects of binary evolution. A spherical geometry is assumed with an inner radius fixed at $R = 0.1$ pc.
The calculations are terminated when the electron fraction reaches 0.01 or when the gas temperature drops below 3000 K. The gas metallicity is set to $0.2\,Z_{\odot}$, while the hydrogen number density and stellar population age are assumed to be $10^{4}$ cm$^{-3}$ and 2 Myr, respectively.
As a result, we find that to change from the field value of $\log(\mathrm{[O\textsc{iii}]\lambda5007/H}\beta) \sim 0.85$ to the stacked value of COSMOS\_ID0\_PC4, $\log(\mathrm{[O\textsc{iii}]\lambda5007/H}\beta) \sim 0.55$, the ionisation parameter $\log U$ needs to change from $-1.8$ to $-2.9$.
Compared to the value $\log U \sim -2$ commonly used for ionisation parameters of high-redshift galaxies \citep[e.g.][]{Grazian2017, Nakajima2018, Saxena2022}, $\log U = -2.9$ is remarkably low.
Therefore, it is difficult to explain the differences discussed below solely by changes in the ionisation parameter.
For the individual-galaxy analysis, we measure gas-phase metallicities for 17 protocluster members and 23 field galaxies with [O\textsc{iii}] detected at S/N $>5$ and with H$\beta$ covered in the observed wavelength range.
The protocluster subsamples contain three, five, seven, and two galaxies for ELAIS-N1\_ID4\_PC1, DEEP2-3\_ID2\_PC2, COSMOS\_ID0\_PC3, and COSMOS\_ID0\_PC4, respectively.
If the flux of the H$\beta$ emission line is less than $1\sigma$, the $1\sigma$ value is used as an upper limit for the H$\beta$ flux.

\begin{figure*}
    \centering
    \includegraphics[width=0.47\linewidth]{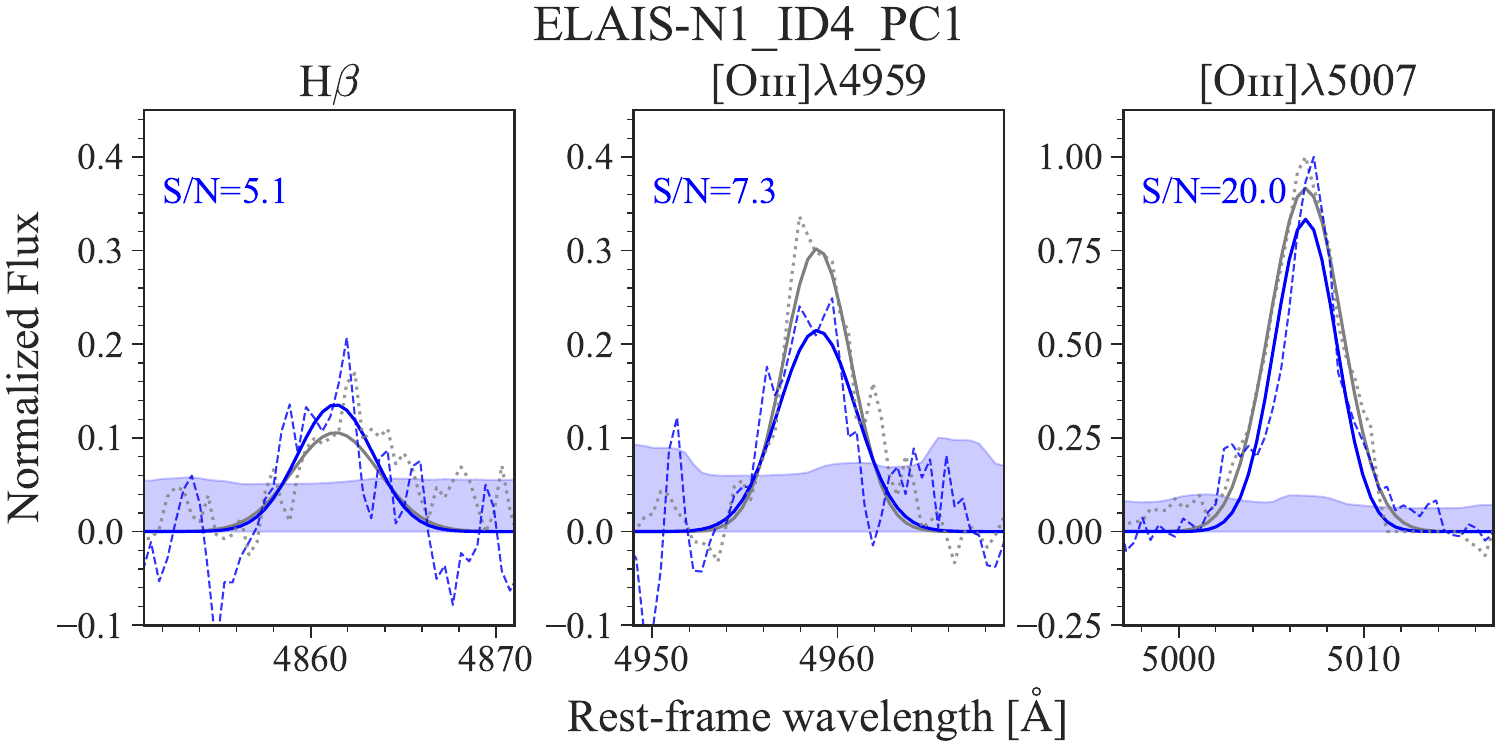}\hspace{0.5cm}
    \includegraphics[width=0.47\linewidth]{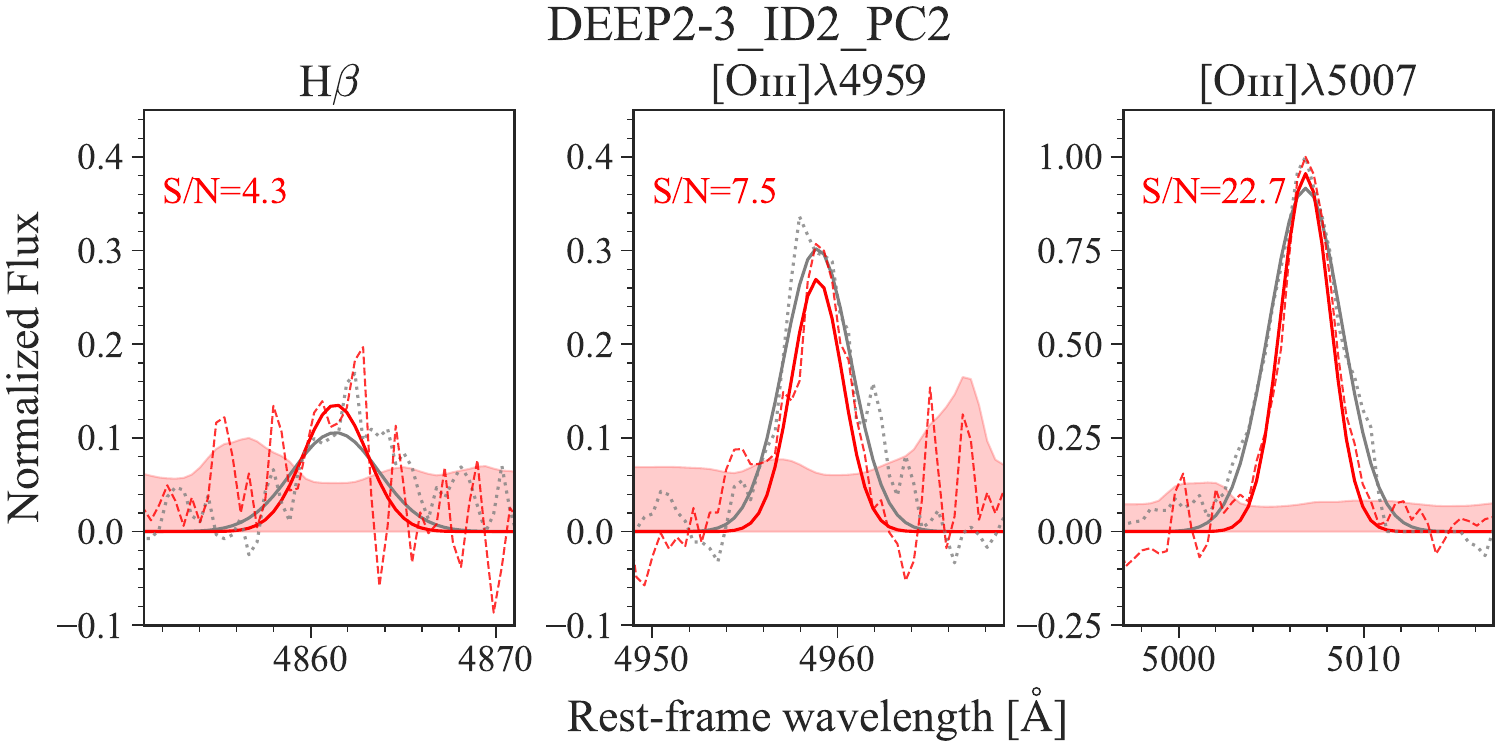}\\\vspace{0.3cm}
    \includegraphics[width=0.47\linewidth]{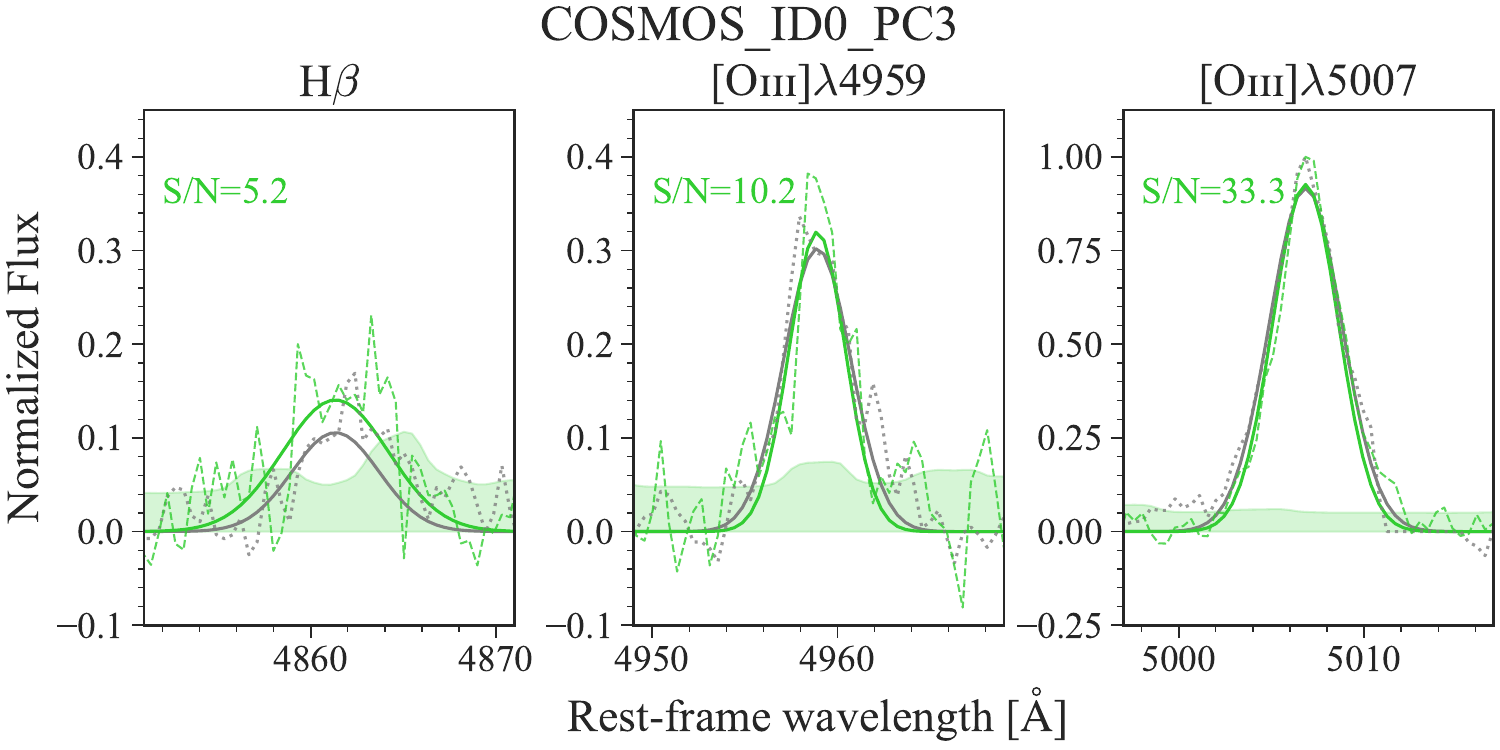}\hspace{0.5cm}
    \includegraphics[width=0.47\linewidth]{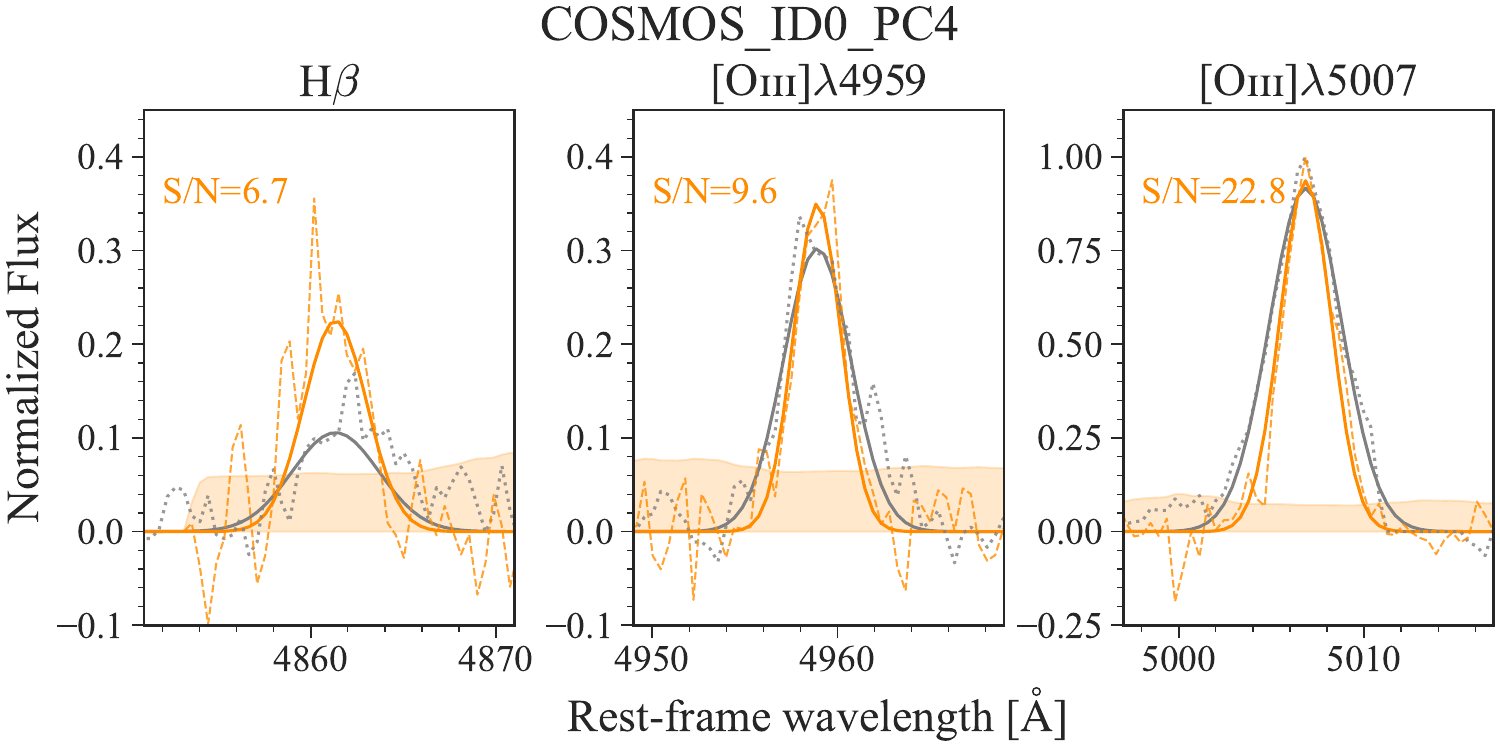}
    \caption{Stacked emission line spectra (left: H$\beta$, middle: [O\textsc{iii}]$\lambda4959$, right: [O\textsc{iii}]$\lambda5007$) of protocluster members in ELAIS-N1\_ID4\_PC1 (top left), DEEP2-3\_ID2\_PC2 (top right), COSMOS\_ID0\_PC3 (bottom left), and COSMOS\_ID0\_PC4 (bottom right). The coloured dashed and solid lines indicate the stacked spectrum and best-fit model of each protocluster, and the shaded region indicates the 1$\sigma$ uncertainty. 
    The grey dotted and solid lines indicate the stacked spectrum and best-fit model of the field galaxies.
    The vertical axis is normalised by the maximum value of the stacked [O\textsc{iii}] spectrum.}
    \label{fig:stack1dspec}
\end{figure*}

\begin{figure*}
    \centering
    \includegraphics[width=\linewidth]{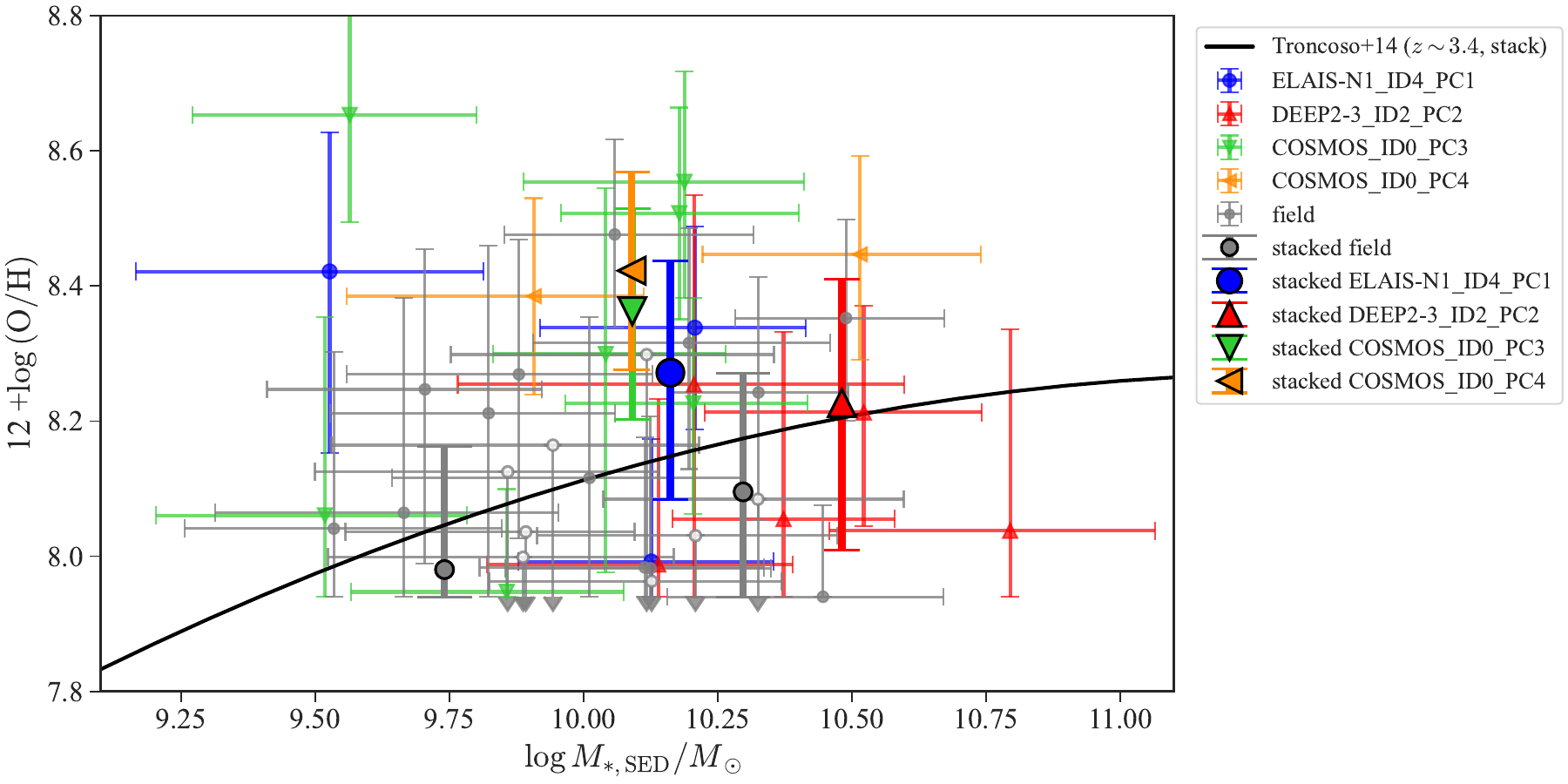}
    \caption{Stellar mass vs. metallicity. The large, filled coloured symbols represent the stacked spectra of protocluster members, while the smaller, lighter coloured symbols show individual galaxy measurements. The colour coding is consistent with Fig.~\ref{fig:ms}. 
    The light grey circles represent field galaxies (i.e., [O\textsc{iii}]-detected non-members).
    The grey line shows the stellar mass-metallicity relation at $z\sim3.5$ from \citet{Troncoso2014} for comparison.}
    \label{fig:metal}
\end{figure*}

Fig.~\ref{fig:stack1dspec} presents the stacked emission line spectra of H$\beta$, [O\textsc{iii}]$\lambda$4959, and [O\textsc{iii}]$\lambda$5007 for each protocluster.
The uncertainties of metallicity are estimated by propagating the flux errors of each spectrum and the uncertainty in the calibration relation \citep{Sanders2025}.
The S/N of H$\beta$ and [O\textsc{iii}] in the stacked spectra are 4.3--6.1 and 20.0--33.3, respectively.
Fig.~\ref{fig:metal} displays the relationship between stellar mass, derived from SED fitting, and gas-phase metallicity, determined from the [O\textsc{iii}]/H$\beta$ emission line ratio.
The stellar mass for the stacked spectra is calculated as the weighted mean of the stellar masses of the individual galaxies, using the S/N of the [O\textsc{iii}]$\lambda5007$ emission line as weights.
The metallicity derived from the stacked spectrum of field galaxies is consistent with the mass-metallicity relation of LBGs at $z\sim3.4$ reported by \citet{Troncoso2014} within the uncertainties.
It should be noted that the mass-metallicity relation can vary depending on the sample selection and the metallicity calibration method used \citep{Sanders2021}.
The stacked spectra show that COSMOS\_ID0\_PC3 and COSMOS\_ID0\_PC4 lie above our field sample and the mass--metallicity relation of \citet{Troncoso2014} at comparable stellar masses, with offsets at the 1.4$\sigma$ and 2.0$\sigma$ levels, respectively. 
In contrast, ELAIS-N1\_ID4\_PC1 and DEEP2-3\_ID2\_PC2 are consistent with the mass--metallicity relation of \citet{Troncoso2014} within the uncertainties.
Although the significance is modest, the data suggest a systematic trend that
the lower-mass protoclusters have marginally higher metallicities. 
However, the difference in the average stellar mass among the four protoclusters is no more than 0.5 dex, which is not particularly large.
In addition, the stellar mass of DEEP2-3\_ID2\_PC2 is highly uncertain because it is estimated from SED fitting without NIR data. 
Furthermore, as shown in Section~\ref{ssec:halomass}, the halo masses of our four protoclusters also agree within 0.4 dex. 
Therefore, it is difficult to draw a strong conclusion that the observed metallicity differences are clearly attributed to differences in stellar mass or halo mass.
We discuss this issue in Sec.\ref {ssec:phys_origin}.
It is important to note that the apparent metallicity differences among the protoclusters are modest, roughly at the $1.4-2.0\sigma$ level, and remain subject to considerable uncertainty. 
Given the overlapping error bars, we cannot claim a statistically significant difference among the protoclusters in the present study. 
The following discussion should therefore be regarded as an exploration of possible interpretations, assuming that these marginal trends reflect underlying physical differences.

It is also important to consider the potential influence of weak Active Galactic Nuclei (AGN) on our metallicity measurements.
Recent studies have reported the presence of AGN in some high-redshift protoclusters \citep[e.g.][]{Chapman2024}. 
If weak AGN are present in some of our target galaxies, their ionising radiation could contribute to the [O\textsc{iii}] emission, potentially leading to high [O\textsc{iii}]/H$\beta$ ratios. 
This, in turn, would result in an underestimation of the gas-phase metallicity when using diagnostics based on these lines.
For example, if some members of the ELAIS-N1\_ID4\_PC1 and DEEP2-3\_ID2\_PC2 protocluster host AGN, their apparent lower metallicities might be, at least in part, an effect of AGN contamination.
It is worth noting that, although the S/N is low, none of the galaxies in our sample exhibit broad H$\beta$ emission, suggesting that the presence of type-I AGN is unlikely.
However, the possibility of type-II AGN cannot be ruled out.
Disentangling the effects of AGN from pure star formation typically requires additional diagnostic tools, such as the [N\textsc{ii}]/H$\alpha$ ratio in combination with [O\textsc{iii}]/H$\beta$ (the BPT diagram; \citealt{Baldwin1981}), or observations of high-ionisation lines. 
Future observations with the James Webb Space Telescope (JWST), capable of detecting H$\alpha$ and [N\textsc{ii}] at $z\sim3.4$, will be crucial to assess the potential impact of AGN on our metallicity estimates and to identify AGN hosts in our sample.

\section{Discussion}
\label{sec:discussion}

\subsection{Comparison with previous studies}
\label{ssec:discussion_previous}

\begin{figure*}
    \centering
    \includegraphics[width=\linewidth]{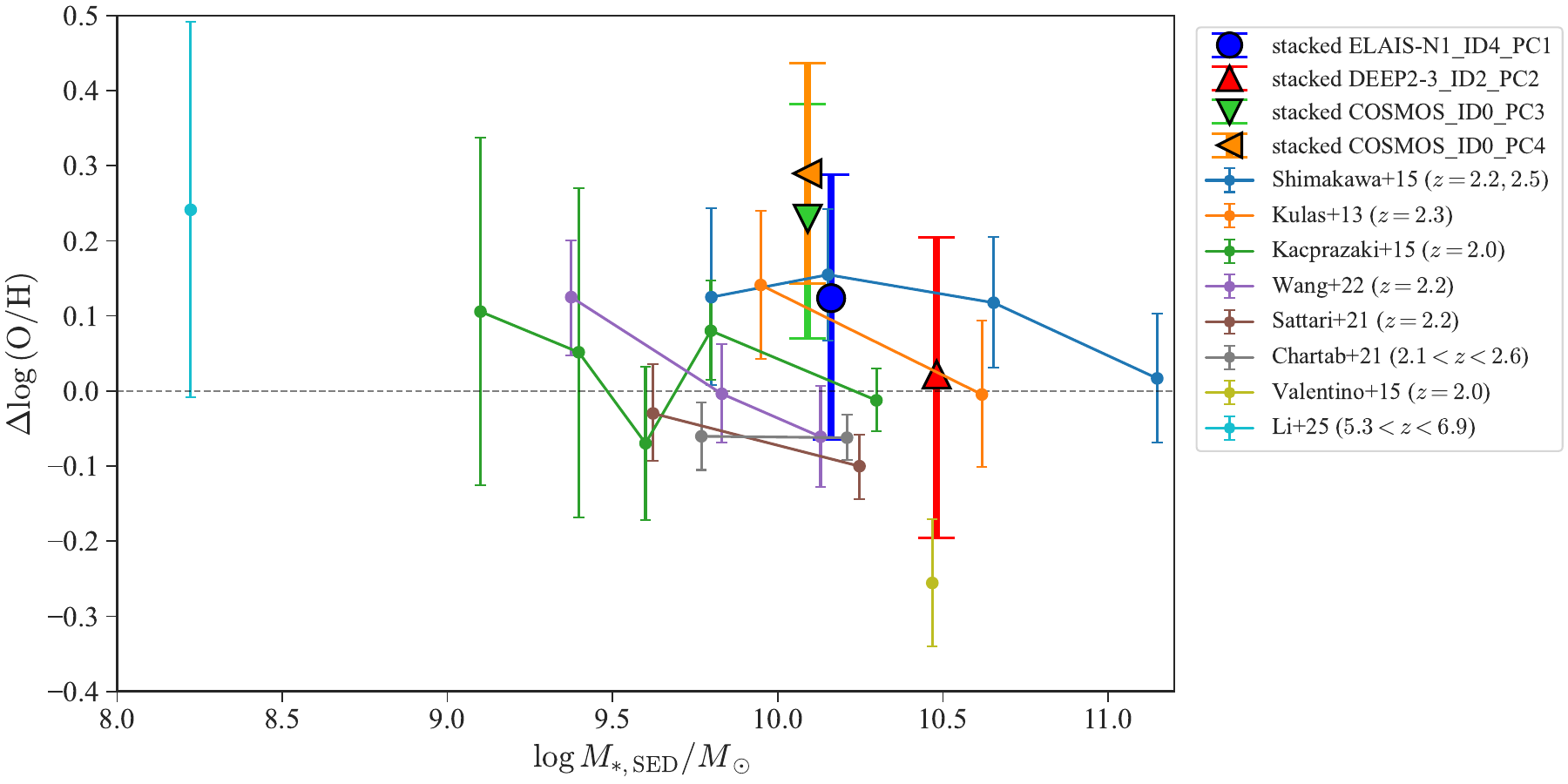}
    \caption{Stellar mass vs. the offset of metallicity from that of field galaxies. 
    The large, filled coloured symbols represent the stacked spectra of protocluster members, while the smaller coloured symbols show the results of previous studies \citep{Shimakawa2015, Kulas2013, Kacprzak2015, Wang2022, Sattari2021, Chartab2021, Valentino2015, Li2025}. The colour coding is consistent with Fig.~\ref{fig:ms}.}
    \label{fig:deltametal}
\end{figure*}

We observe a possible diversity in the gas-phase metallicities within our protoclusters at $z\sim3.4$; these are marginally higher in some systems than the field, while in others it is comparable (Section~\ref{ssec:metal}). 
However, these differences are within or close to the uncertainties and do not provide statistical evidence for distinct metallicity populations.
Fig.~\ref{fig:deltametal} shows the stellar mass versus the metallicity offset from the field mass-metallicity relation by \citet{Troncoso2014} for our protocluster members, along with results from previous studies at $z\sim2$ \citep{Shimakawa2015, Kulas2013, Wang2022, Kacprzak2015, Valentino2015,Sattari2021,Chartab2021} and $z>5$ \citep{Li2025}.
Previous studies of protoclusters at $z \gtrsim 2$ have reported a range of metallicities for member galaxies relative to field galaxies.
Some studies find that protocluster members tend to have higher metallicities \citep{Kulas2013, Wang2025}, especially at the low-mass end \citep[e.g.][]{Shimakawa2015, Li2025}, while others report lower metallicities in protoclusters \citep[e.g.][]{Valentino2015, Chartab2021, Sattari2021, Zhou2025}. 
Still others find that protocluster members have metallicities comparable to field galaxies, or a mix of higher and lower values \citep[e.g.][]{Kacprzak2015, Wang2022}.
Our results, which show diverse metallicities among the four protoclusters, indicate that this diversity extends to $z\sim3.4$ in the similar mass range as previous studies, implying that the variety of metallicities observed in protoclusters may be an intrinsic characteristic.

We also compare our findings with recent studies on metallicity in even higher redshift protoclusters.
Observational efforts are pushing the study of chemical enrichment in the early Universe, with some works suggesting that environmental influences on metallicity are present even at $z>3$. 
For example, \citet{Li2025} compiled data for protoclusters at $z<5$ and argued that their members generally exhibit higher metallicities than field galaxies. 
At $z \sim 7.9$, \citet{Witten2025} reported evidence for elevated metallicities in a spectroscopically confirmed protocluster, potentially indicating early chemical enrichment in dense environments. 
Furthermore, observations of a $z \sim 7.88$ protocluster by \citet{Morishita2025} revealed significant scatter in metallicity among member galaxies. 
While direct comparison is difficult because the target galaxy populations are different, combining these pioneering studies at high redshifts with the current results for protoclusters at $z\sim3.4$ hints that diverse evolutionary pathways and varied chemical enrichment histories might be a persistent feature of galaxy assembly in overdense regions throughout cosmic time.
Our results may be broadly consistent with the possibility that chemically diverse protocluster environments already exist by $z\sim3.4$.

Theoretical studies and simulations also present a variety of predictions in the metallicity of protocluster galaxies at $z>3$.
For instance, the FOREVER22 simulation \citep{Yajima2022} indicates that protocluster regions with halo masses around $M_\mathrm{halo} \sim 10^{13} \, \mathrm{M}_\odot$ at $z=3$ have metallicities comparable to the mean field.
In contrast, \citet{Rowntree2025}, using the Horizon Run 5 cosmological hydrodynamical simulation, suggest that at $z=3.5$, the metallicity of protocluster galaxies (in halos with $M_\mathrm{halo} > 10^{12.16} \, \mathrm{M}_\odot$) starts to deviate from field galaxies, particularly for galaxies with stellar masses $M_* < 10^{9.8} \, \mathrm{M}_\odot$, where protocluster members tend to be more metal-rich.
Exploring the reasons behind these differing results may provide insights into the physical mechanisms driving diverse metal enrichment in protoclusters.

\subsection{Physical mechanisms driving diverse metallicities in protoclusters at $z \sim 3.4$}
\label{ssec:phys_origin}

As we have seen, our sample exhibits intriguing trends in metallicity although the uncertainties are still large.
In this section, we explore the potential physical origins of these differences, assuming they are real.
First, we consider the interpretation of the apparent metallicity enrichment in two of our protoclusters: COSMOS\_ID0\_PC3 and COSMOS\_ID0\_PC4.
One possible interpretation of the marginal metallicity enhancement seen in COSMOS\_ID0\_PC3 and COSMOS\_ID0\_PC4 is more efficient recycling of metal-rich outflows in overdense environments.
If real, such a trend would be broadly consistent with scenarios in which gravitational confinement and the surrounding medium help galaxies retain enriched gas.
This interpretation is supported by simulations \citep{Oppenheimer2008, Dave2011} and some previous observational studies \citep[e.g.][]{Kulas2013, Shimakawa2015}.
Since our protocluster member galaxies and field galaxies are both selected as LBGs and  [O\textsc{iii}] emitters, the difference in metallicity between the two is minimally affected by sample selection bias, as noted in other studies \citep{Shimakawa2015,Chartab2021, Sattari2021}.
Additionally, the stripping of gas reservoirs due to accretion onto a single halo, as discussed by \citet{Shimakawa2015}, is less likely to occur in unvirialised systems at $z\sim3.4$.
Therefore, if the observed metallicity offsets are real, recycling of enriched gas within the protocluster environment may contribute to the relatively high metallicities seen in some systems.

On the other hand, DEEP2-3\_ID2\_PC2 have metallicities equivalent to that of field galaxies at a similar stellar mass, and the metallicity of ELAIS-N1\_ID4\_PC1 is also comparable to that of field galaxies within the margin of uncertainty. If the modest differences among the protoclusters are real, ELAIS-N1\_ID4\_PC1 might be in an intermediate state between DEEP2-3\_ID2\_PC2 and COSMOS\_ID0\_PC3,4.
Previous studies at $z\sim2$ have also reported metal deficiency in protoclusters, suggesting a mechanism whereby metallicity in protoclusters can be diluted by metal-poor cold gas inflow
\citep{Valentino2015,Chartab2021,Sattari2021}.
As shown in Fig.~\ref{fig:halo_mass_evolution}, all our protoclusters are located in the cold-in-hot regime based on the theory of \citet{Dekel2006}, indicating that cold gas inflow can exist in these environments.

If the diversity in metallicity across the four protoclusters is real, it would be reasonable to consider that the inflow effect might be stronger in DEEP2-3\_ID2\_PC2 and ELAIS-N1\_ID4\_PC1, which lack apparent metallicity enhancement, while it is weaker in COSMOS\_ID0\_PC3,4, which show metallicity enhancement.
However, we again stress that these apparent metallicity differences are only at modest significance in the present study. 
Therefore, the possible role of metal-poor inflow discussed here should be regarded as tentative.

Then, why does the effect of inflow differ among protoclusters?
The four protoclusters exhibit comparable halo masses (Section~\ref{ssec:halomass}), and no significant differences in stellar mass or SFR (Section~\ref{ssec:sedfit_properties}) are observed between the protoclusters with apparently higher metallicities and those whose metallicities are consistent with the field. 
This implies that variations in gas inflow strength inferred from these modest metallicity differences might not be primarily driven by the intrinsic properties of the protoclusters themselves but rather by the surrounding environmental conditions.
\citet{Yajima2022} investigated protoclusters with halo masses above $10^{13} \, \mathrm{M}_\odot$ at $z\sim3$ in the hydrodynamical simulations and found various inflow patterns, including some protoclusters connected to several large filaments and others formed at the junction of finer filaments.
Relatively low-mass protoclusters are expected to reside not at the nodes where filaments converge but rather in their surrounding regions, where the morphology of the large-scale structure is more complex.
Such peripheral locations may naturally produce the diversity in gas supply observed among our protoclusters with halo masses of $\sim 10^{12.5} \, \mathrm{M}_\odot$.
The diversity of environments around protoclusters is also supported by the varied halo mass growth histories of overdensities \citep[e.g.][]{Muldrew2015}.
These findings emphasise the necessity of understanding the chemical properties of galaxies within protoclusters not merely as a function of mass, but also by considering the diversity of the large-scale filamentary structures connecting them.
A recent report by \citet{Yang2026} concludes that whether the mass-metallicity relation of overdense regions lies above or below that of the field depends on the stellar mass of the member galaxy. 
While they combined three overdense regions at $z\sim2$ to derive the mass-metallicity relation, it should be noted that, as discussed above, the situation may differ for individual protoclusters.

In summary, if the observed metallicity differences are real despite the overlapping uncertainties, recycling in protoclusters could lead to increased metallicities, while the dilution effect from metal-poor inflows might also be present depending on the surrounding environment of the protocluster. 
The balance between these two effects may determine whether the metallicity of the protocluster appears to be greater or less than that of the field.

\section{Summary and Conclusions}
\label{sec:conclusion}
We conduct spectroscopic follow-up observations of protocluster candidates at
$z\sim3.4$, originally identified using the PCFNet machine learning algorithm from HSC imaging data. Our NIR spectroscopy successfully identifies four protoclusters. We analyse the physical properties of their member galaxies through SED fitting and emission line diagnostics. Our main findings are summarised as follows:

\begin{enumerate}
    \item 
    We identify 57 [O\textsc{iii}] emitters and three Pa$\alpha$ emitters. Among the 57 [O\textsc{iii}] emitters, 17 additionally show significant ($>3\sigma$) H$\beta$ detections. 
    We confirm four protoclusters at $z=3.316,\ 3.403,\ 3.408$ and $3.425$ in three of the four targeted regions, while the remaining region (DEEP2-3\_ID1) does not show a significant redshift peak by the present observation.
    These protoclusters contain five, five, eight, and five member galaxies for ELAIS-N1\_ID4\_PC1, DEEP2-3\_ID2\_PC2, COSMOS\_ID0\_PC3, and COSMOS\_ID0\_PC4, respectively, which are more than five times higher density peaks than a uniform distribution.
    \item The detection rate of protocluster members among the galaxies predicted by PCFNet to have $\sigma_\mathrm{prob}>2.5$ is 40\%, which is consistent with the precision predicted by PCFNet and notably higher than the rates achieved by conservative methods. 
    By candidate regions, three of the four targeted candidate regions are confirmed, corresponding to an observed protocluster detection purity of 75\%, consistent with the purity predicted by PCFNet.
    \item Protocluster member galaxies predominantly reside on the star-forming main sequence. 
    Although some systems have slightly higher mean stellar masses than the field, the difference is not statistically significant. 
    Given the small number of spectroscopically confirmed members in each system, our data do not provide strong evidence for a systematic stellar-mass excess.
    \item We estimate the mass of the most massive halo in each protocluster from the most massive stellar mass of the spectroscopically confirmed member using the SHMR, yielding $\log{M_\mathrm{halo}/\mathrm{M_\odot}}=12.4^{+0.38}_{-0.29}, 12.8^{+0.49}_{-0.31}, 12.4^{+0.39}_{-0.26}$ and $12.6^{+0.41}_{-0.28}$ for ELAIS-N1\_ID4\_PC1, DEEP2-3\_ID2\_PC2, COSMOS\_ID0\_PC3, and COSMOS\_ID0\_PC4, respectively.
    A comparison with cosmological simulations suggests that these main halos are consistent with the progenitors of relatively low-mass clusters.
    \item We estimate the total SFRs and volumes of the protoclusters using the VMC approach based on the protocluster membership probabilities derived by PCFNet. 
    The derived SFRDs are substantially higher than the cosmic SFRD at this redshift and are comparable to those reported for other $z\sim2\text{--}4$ protoclusters, although caution is required when 
    comparing them with literature due to differences in assumed geometry, volume definition, and methodology.
    \item Analysis of gas-phase metallicity suggests that COSMOS\_ID0\_PC3 and COSMOS\_ID0\_PC4 may have marginally higher average metallicities than field galaxies with similar stellar masses, whereas ELAIS-N1\_ID4\_PC1 and DEEP2-3\_ID2\_PC2 are broadly consistent with the field. However, these differences are only at the $\sim1.4-2.0\sigma$ level and are not statistically significant in the present sample.
\end{enumerate}

These results underscore the importance of targeted spectroscopic surveys of machine-learning-selected candidates for efficiently identifying and characterising protoclusters at high redshifts. 
Our findings provide new insights into the diversity of chemical properties within these early large-scale structures and their role in shaping galaxy evolution in the early Universe. 
Future observations with facilities such as JWST will be crucial for further dissecting the detailed physical processes at play within these fascinating cosmic laboratories.
Also, PCFNet can be applied to next-generation wide surveys such as LSST and Roman, and more protoclusters are expected to be detected in the future.

\section*{Acknowledgements}

We thank Kazuhiro Shimasaku, Ichi Tanaka, Tadayuki Kodama, Ken Mawatari, Akito Kusaka, Kent Fujiwara and Makoto Ando for their useful discussion.
Y.T. was supported by Forefront Physics and Mathematics Program to Drive Transformation (FoPM), a World-leading Innovative Graduate Study (WINGS) Program, the University of Tokyo, and JSPS KAKENHI Grant Number JP23KJ0726.
N.K. was supported by the Japan Society for the Promotion of Science through Grant-in-Aid for Scientific Research 25H00663, 25K01038, 25K01044.
The authors wish to express their gratitude to GitHub Copilot, Claude Sonnet 4, and Gemini 2.5 Pro for providing valuable assistance in the translation and linguistic refinement of this article. The authors remain solely responsible for the final content of the paper.

The Hyper Suprime-Cam (HSC) collaboration includes the astronomical communities of Japan and Taiwan, and Princeton University. The HSC instrumentation and software were developed by the National Astronomical Observatory of Japan (NAOJ), the Kavli Institute for the Physics and Mathematics of the Universe (Kavli IPMU), the University of Tokyo, the High Energy Accelerator Research Organization (KEK), the Academia Sinica Institute for Astronomy and Astrophysics in Taiwan (ASIAA), and Princeton University. Funding was contributed by the FIRST program from the Japanese Cabinet Office, the Ministry of Education, Culture, Sports, Science and Technology (MEXT), the Japan Society for the Promotion of Science (JSPS), Japan Science and Technology Agency (JST), the Toray Science Foundation, NAOJ, Kavli IPMU, KEK, ASIAA, and Princeton University. 

This paper makes use of software developed for Vera C. Rubin Observatory. We thank the Rubin Observatory for making their code available as free software at \url{http://pipelines.lsst.io/}.

This paper is based on data collected at the Subaru Telescope and retrieved from the HSC data archive system, which is operated by the Subaru Telescope and Astronomy Data Center (ADC) at NAOJ. Data analysis was in part carried out with the cooperation of Center for Computational Astrophysics (CfCA), NAOJ. We are honored and grateful for the opportunity of observing the Universe from Maunakea, which has the cultural, historical and natural significance in Hawaii. 

The Pan-STARRS1 Surveys (PS1) and the PS1 public science archive have been made possible through contributions by the Institute for Astronomy, the University of Hawaii, the Pan-STARRS Project Office, the Max Planck Society and its participating institutes, the Max Planck Institute for Astronomy, Heidelberg, and the Max Planck Institute for Extraterrestrial Physics, Garching, The Johns Hopkins University, Durham University, the University of Edinburgh, the Queen’s University Belfast, the Harvard-Smithsonian Center for Astrophysics, the Las Cumbres Observatory Global Telescope Network Incorporated, the National Central University of Taiwan, the Space Telescope Science Institute, the National Aeronautics and Space Administration under grant No. NNX08AR22G issued through the Planetary Science Division of the NASA Science Mission Directorate, the National Science Foundation grant No. AST-1238877, the University of Maryland, Eotvos Lorand University (ELTE), the Los Alamos National Laboratory, and the Gordon and Betty Moore Foundation.

This work made use of v2.2.1 of the Binary Population and Spectral Synthesis (BPASS) models as described in \cite{Eldridge2017} and \cite{Stanway2018}.

\section*{Data availability}
The source code of PCFNet is available in the GitHub page at \url{https://github.com/YoshihiroTakeda/PCFNet}. Our spectroscopic data can be shared on a reasonable request.



\bibliographystyle{mnras}
\bibliography{main} 

\appendix

\section{Impact of different Halo Mass Estimation Methods}
\label{sec:halomassestimation}

\begin{figure}
    \centering
    \includegraphics[width=\linewidth]{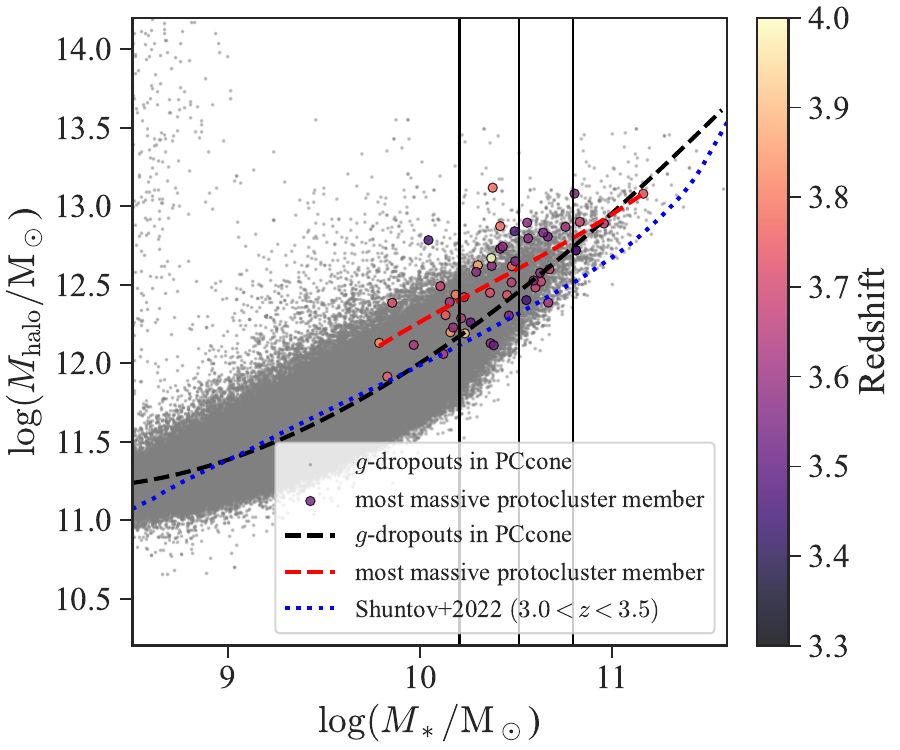}
    \caption{Stellar mass vs. halo mass relation for $g$-dropout galaxies in PCcone simulation.
    The coloured points indicate the galaxies with the highest stellar mass in the galaxy groups predicted as protoclusters by PCFNet.
    The lines represent the best-fit relations for all $g$-dropout galaxies (black dashed), the galaxies with the highest stellar mass in the galaxy groups predicted as protoclusters by PCFNet (red), and the relation from \citet{Shuntov2022} (blue).}
    \label{fig:halomassestimation}
\end{figure}

Fig.~\ref{fig:halomassestimation} shows the relation between stellar mass and halo mass for $g$-dropout galaxies selected from the PCcone simulation.
The figure includes best-fit lines for all $g$-dropout galaxies (black), for the galaxies with the highest stellar mass in the galaxy groups predicted as protoclusters by PCFNet (red), and the relation from \citet{Shuntov2022} (blue).
These lines are consistent within $\sim0.2$ dex, indicating that different methods of estimating halo mass based on stellar-to-halo mass ratios do not significantly impact the halo mass estimates.

\section{Comparison of protocluster member, field galaxies and non-detected galaxies}
\label{sec:detect_nondetect}
\begin{figure}
    \centering
    \includegraphics[width=\linewidth]{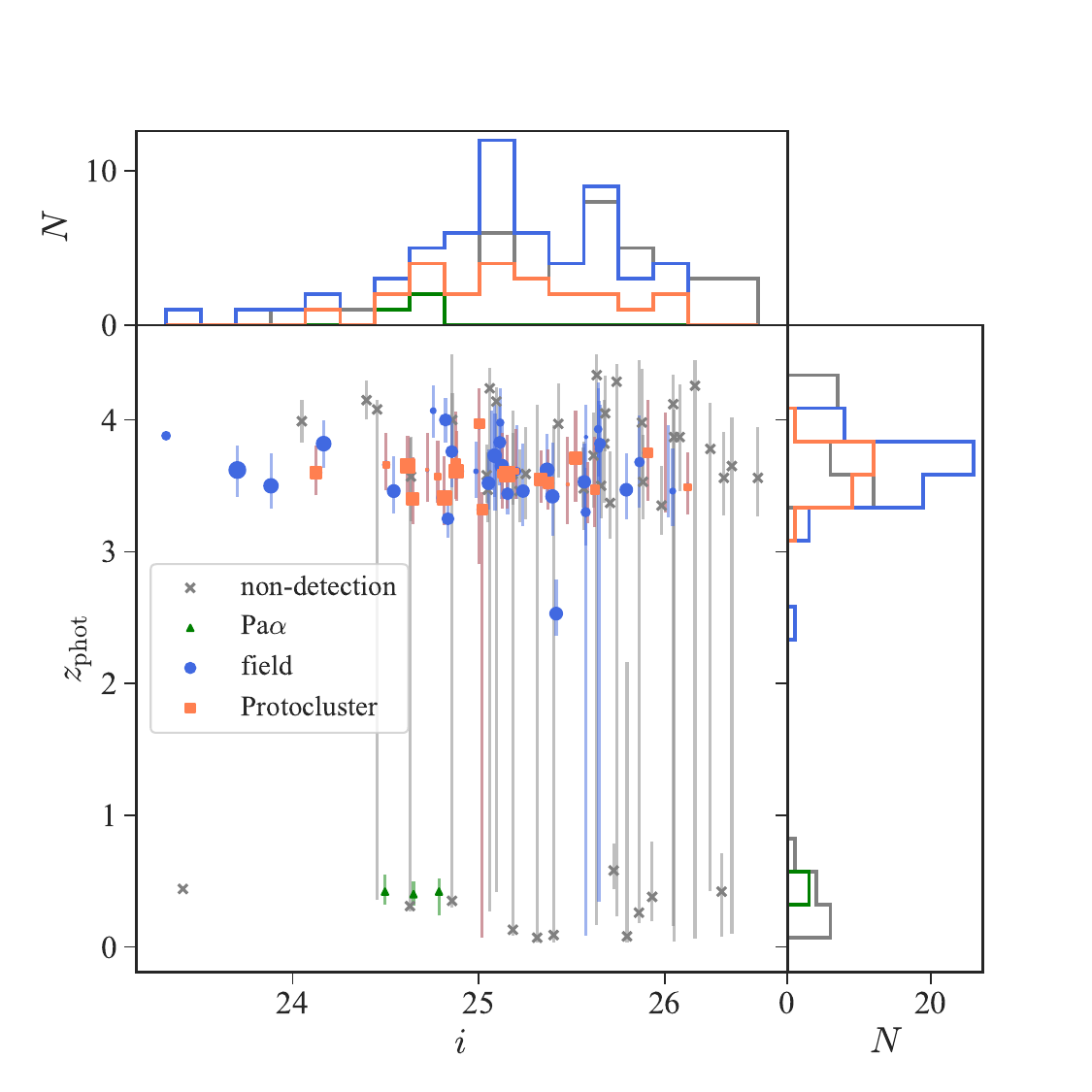}
    \caption{$i$ band magnitude vs. photo-$z$ for the spectroscopically detected field galaxies (blue circle), protocluster members (orange square),  Pa$\alpha$ detected galaxies (green triangle) and non-detected galaxies (grey cross). The size of each marker is scaled with the flux of [O\textsc{iii}]$\lambda5007$ emission line for the detected galaxies.
    The marginal histograms along the horizontal and vertical axes show the distributions of photo-$z$ and $i$-band magnitude, respectively, for each subsample.
    Note that even objects with lower photometric redshifts are selected as spectroscopic targets if their $\sigma_{\mathrm{prob}}>1.5$.
    }
    \label{fig:detect_nondetect}
\end{figure}

\begin{figure}
    \centering
    \includegraphics[width=\linewidth]{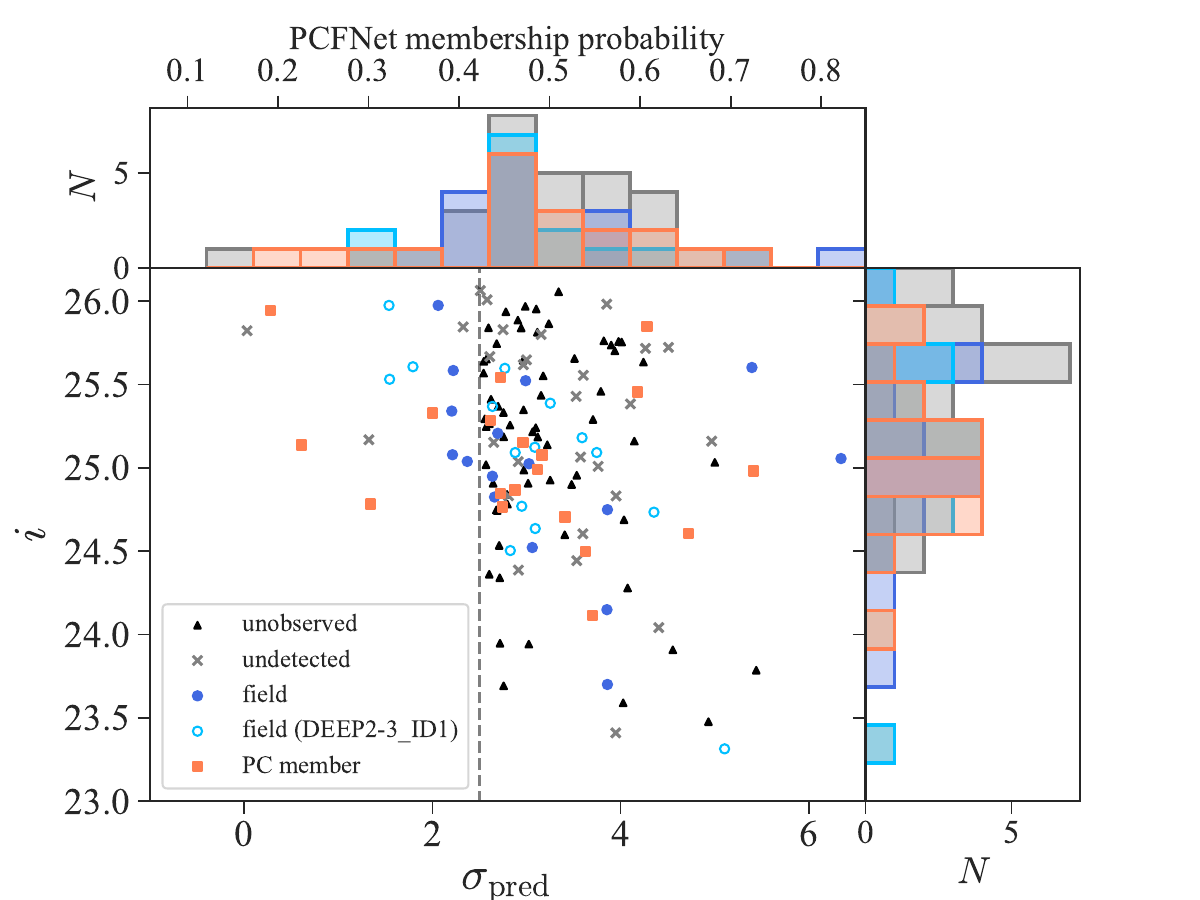}
    \includegraphics[width=\linewidth]{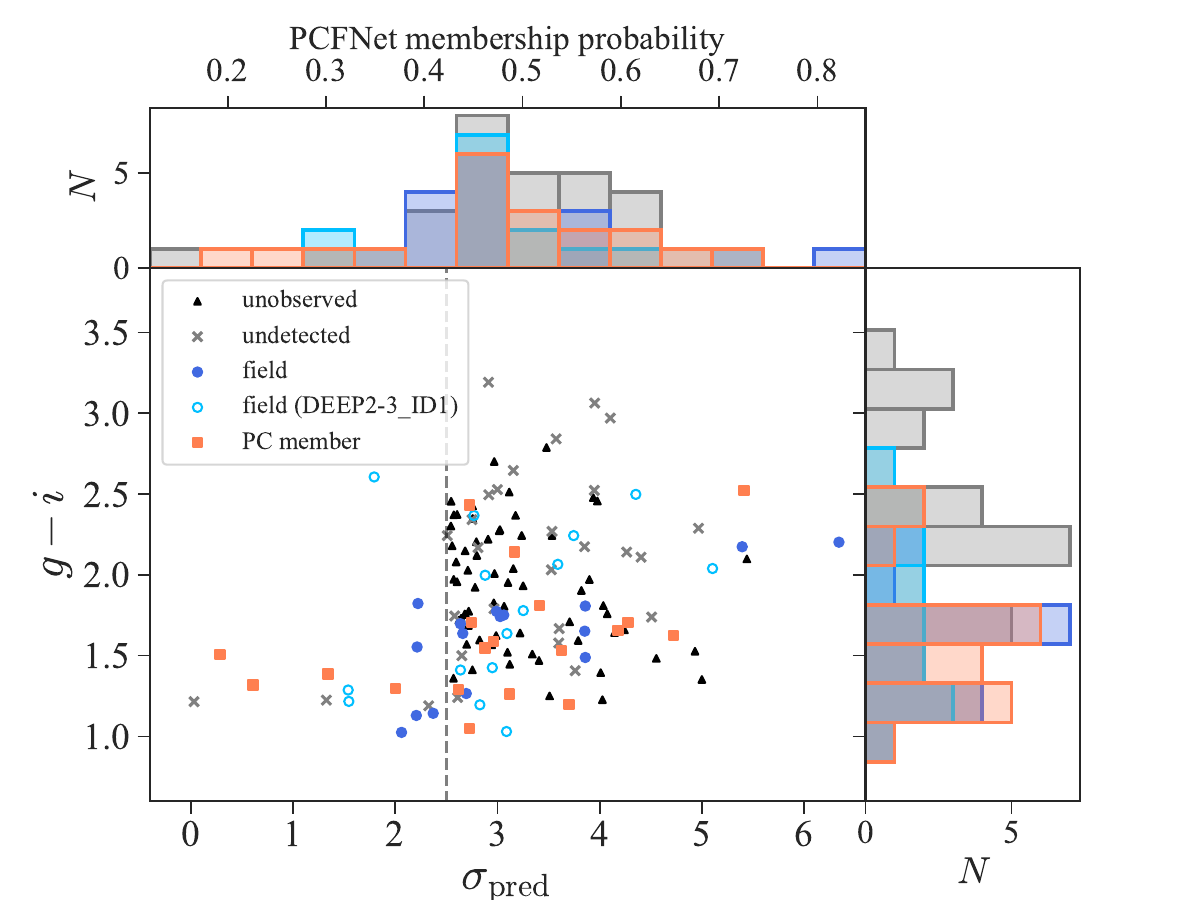}
    \caption{$\sigma_\mathrm{prob}$ vs. $i$ band magnitude (top) and $g-i$ colour (bottom) for the spectroscopically detected field galaxies (blue circle), protocluster members (orange square), undetected galaxies (grey cross), and unobserved galaxies (black triangle). The field galaxies in DEEP2-3\_ID1 are indicated as open circles. 
    The marginal histograms show the distributions of $\sigma_\mathrm{prob}$ and $i$-band magnitude (top), and of $\sigma_\mathrm{prob}$ and $g-i$ colour (bottom) for each subsample, excluding the unobserved galaxies.
    }
    \label{fig:sigma_pred}
\end{figure}

We show the comparison of the $i$ band magnitude and photo-$z$ derived from the joint catalogue between the spectroscopically detected and non-detected galaxies in Fig.~\ref{fig:detect_nondetect}.
The non-detected galaxies tend to have fainter $i$ band magnitudes or photo-$z$ values outside the observable range $2.8<z<3.8$.
We also show the comparison of $\sigma_\mathrm{prob}$ derived from PCFNet between the spectroscopically detected and non-detected galaxies in Fig.~\ref{fig:sigma_pred}.
Similarly, in these figures, within the range $2.5<\sigma_{\mathrm{prob}}<4$, brighter objects are more likely to be spectroscopically detected, and those with $g-i<2$, i.e., expected lower photometric redshifts due to less IGM absorption, are also more likely to be spectroscopically detected.
It should be noted that the PCFNet membership probabilities are not determined only by individual magnitude and colours, and galaxies not classified as member candidates still have a non-zero probability of being protocluster members.
Furthermore, even galaxies with a high probability may remain unobserved due to slit collisions, while conversely, only a very small number of galaxies with a low probability are allocated to the slit. 
Because the spectroscopic observations are incomplete, it is difficult to evaluate the performance of PCFNet based on these results.
Although some galaxies with $\sigma_\mathrm{prob}>4$ are detected as protocluster members despite being faint, the sample size is insufficient to draw definitive conclusions about these trends.
Note that galaxies with $\sigma_\mathrm{prob}<2.5$ are selected as spectroscopic targets with the aid of photo-$z$ information.

\section{Catalogue of the spectroscopically identified galaxies}
Table~\ref{tab:obsemission} lists the properties of the spectroscopically identified galaxies and Fig.~\ref{fig:all_2d_1d} shows their 2D and 1D spectra.

\begin{landscape}
\begin{table}
	\centering
	\caption{
        Properties of the spectroscopically identified galaxies.
        The detection criterion for each emission line is S/N$>3$.
        }
	\label{tab:obsemission}
    \input{tbl/oiii_detected_galaxies_part1}
\end{table}
\end{landscape}
\begin{landscape}
\begin{table}
    \continuedfloat
	\centering
	\caption{
        Continued.
        }
    \input{tbl/oiii_detected_galaxies_part2}
\end{table}
\end{landscape}

\begin{figure*}
    \centering
    \includegraphics[width=\linewidth]{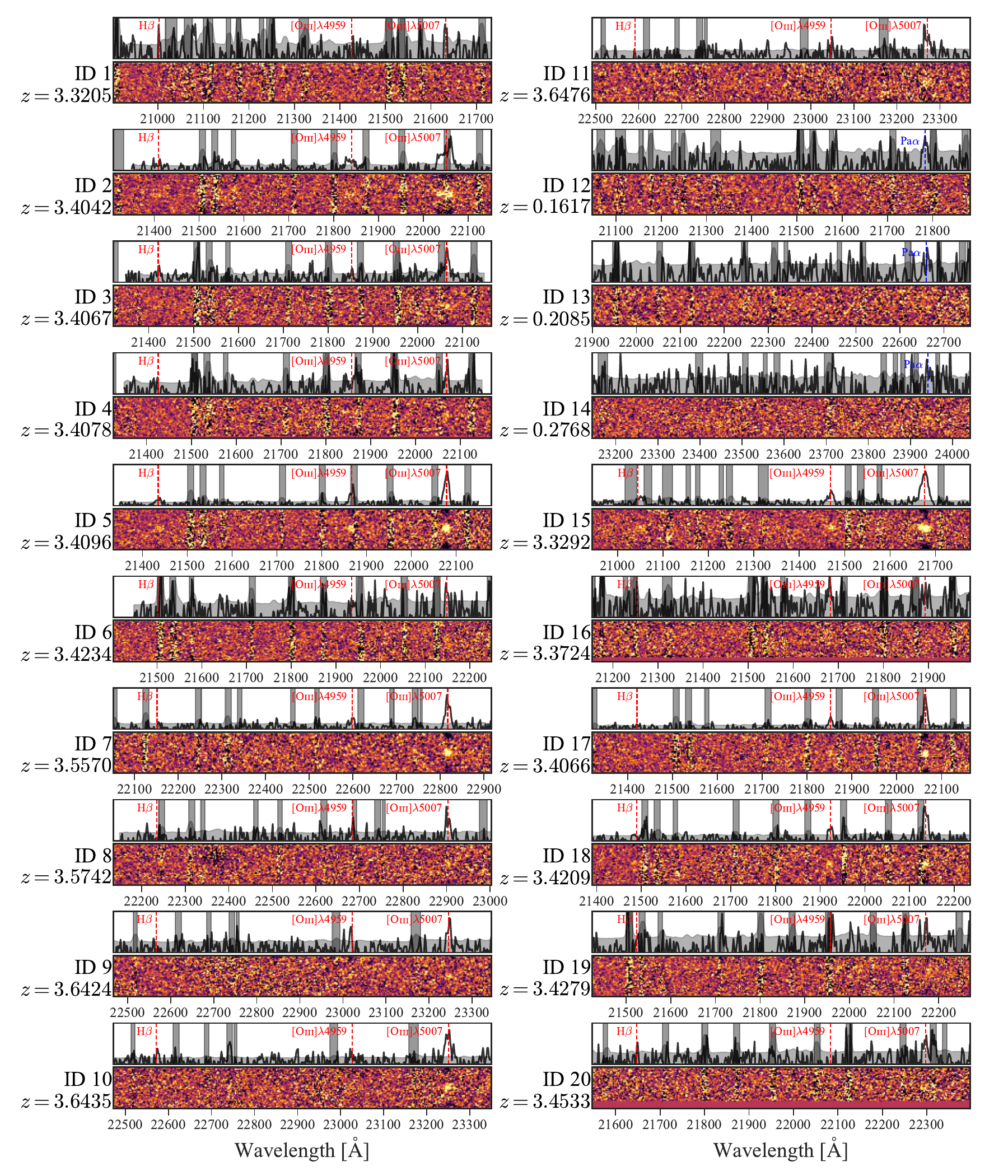}
    \caption{2D and 1D spectra of the spectroscopically identified galaxies. The top panels show individual 1D spectra extracted from the 2D spectrum. The bottom panels show a 2D spectrum.
    Key emission lines, including H$\beta$ and [O\textsc{iii}]$\lambda\lambda$4959, 5007, are labelled.
    Shaded regions in 1D spectra are $1\sigma$ uncertainty in each wavelength grid.
    Vertical dark shaded regions in the 1D spectrum represent the wavelength range of strong sky emission lines.}
    \label{fig:all_2d_1d}
\end{figure*}

\begin{figure*}
    \centering
    \continuedfloat
    \includegraphics[width=\linewidth]{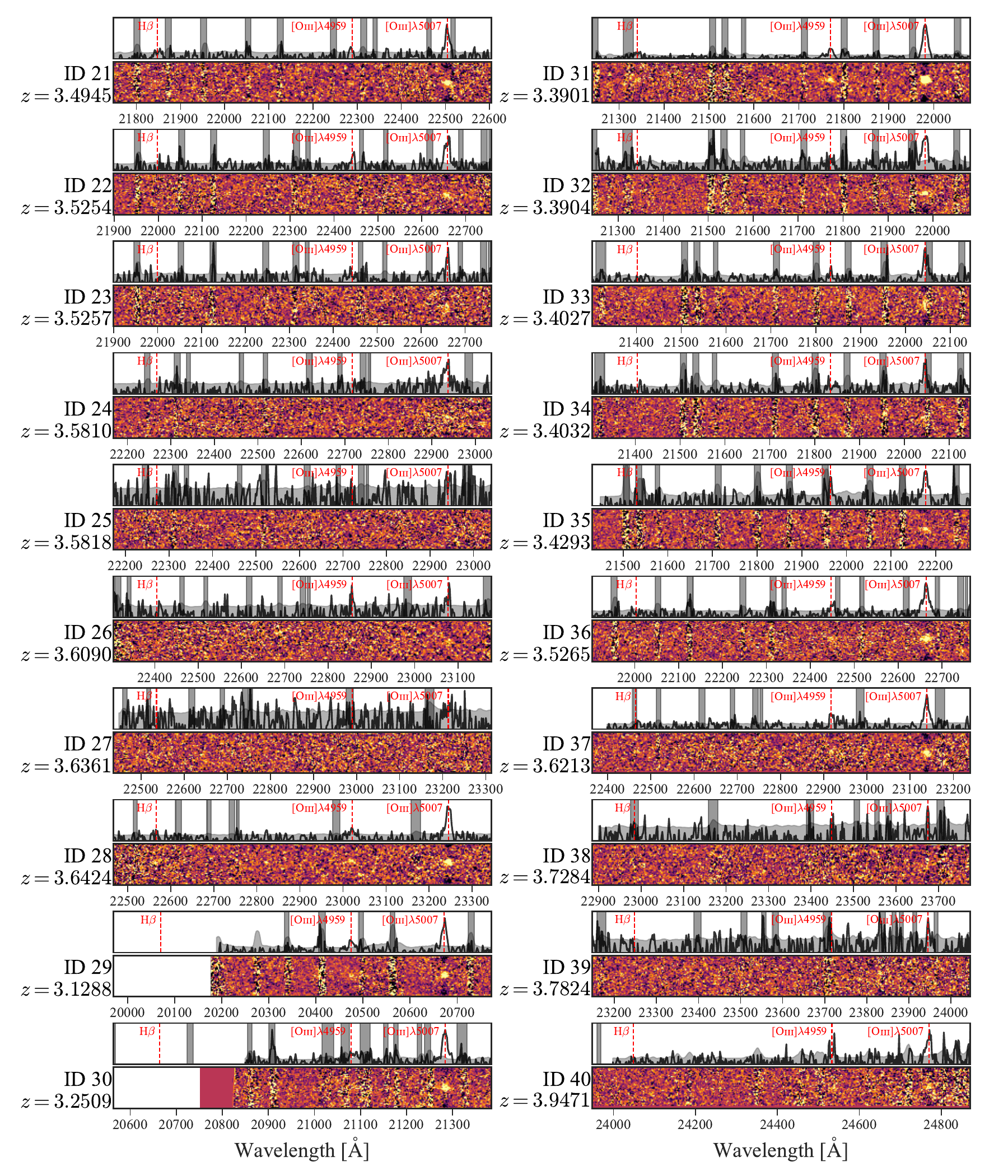}
    \caption{Continued.}
\end{figure*}

\begin{figure*}
    \centering
    \continuedfloat
    \includegraphics[width=\linewidth]{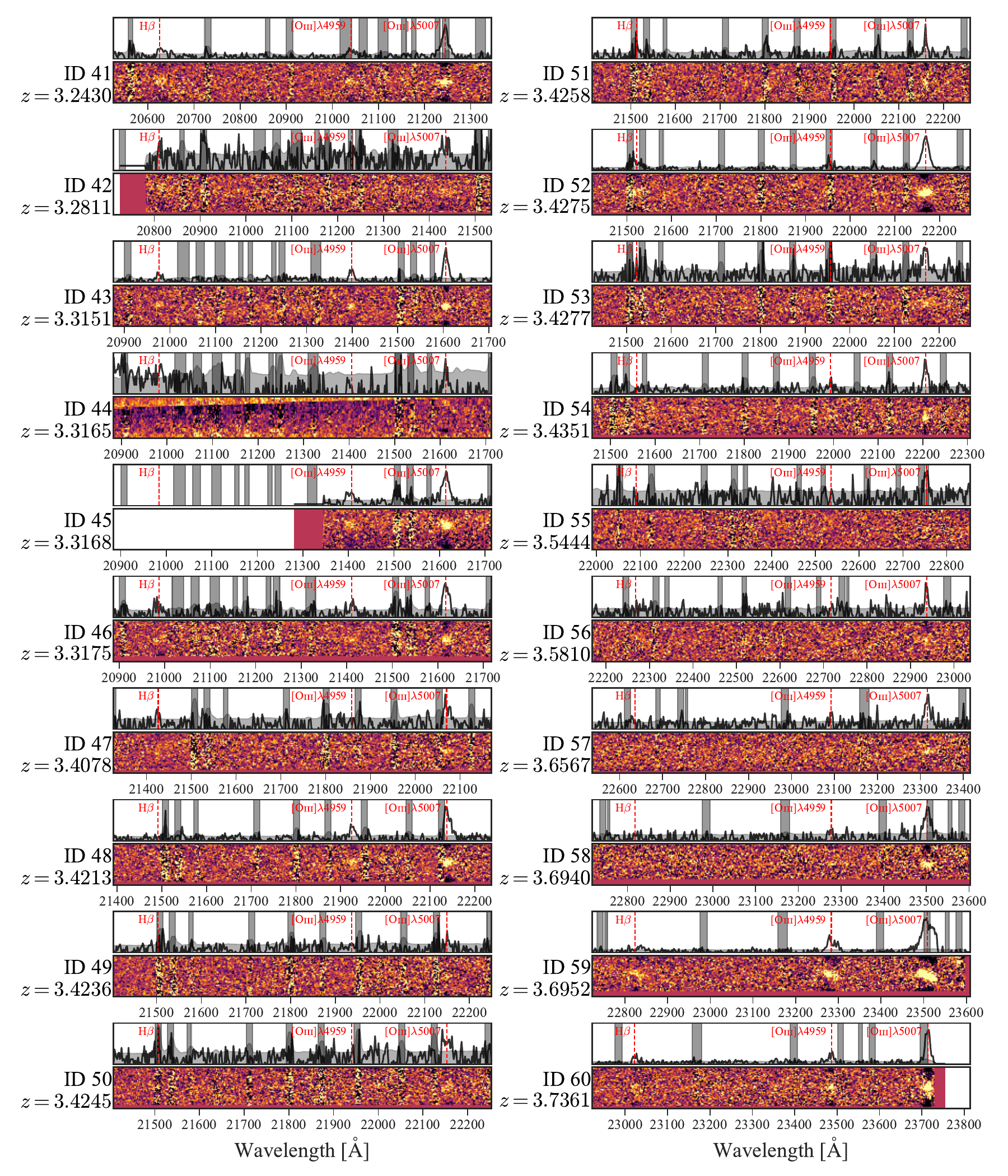}
    \caption{Continued.}
\end{figure*}

\bsp	
\label{lastpage}
\end{document}

%% file: tbl/oiii_detected_galaxies_part1.tex
\begin{tabular}{rcccccccccc}
\toprule
 ID &        Region & R.A. (J2000) & Decl. (J2000) & $z_\mathrm{spec}$ & H$\beta$ Detection & \begin{tabular}{c}[O\textsc{iii}]5007 Flux\\ (10$^{-17}$ erg/s/cm$^2$)\end{tabular} &  $\log{M_*/M_\odot}$ & \begin{tabular}{c}SFR\\ (M$_\odot$/yr)\end{tabular} & \begin{tabular}{c}$A_V$\\ (mag)\end{tabular} & \begin{tabular}{c}Age\\ (Gyr)\end{tabular} \\
\midrule
  1 & ELAIS-N1\_ID4 &  16:12:25.39 &  +55:57:42.66 & $3.3205\pm0.0001$ &                    &                                        $1.0\pm0.3$ & $10.4^{+0.2}_{-0.3}$ &                                   $99^{+28}_{-16}$ &                       $1.31^{+0.13}_{-0.09}$ &                     $0.20^{+0.20}_{-0.11}$ \\
  2 & ELAIS-N1\_ID4 &  16:12:32.14 &  +56:01:25.08 & $3.4042\pm0.0002$ &       $\checkmark$ &                                       $11.1\pm0.8$ & $10.1^{+0.2}_{-0.2}$ &                                   $40^{+15}_{-10}$ &                       $0.93^{+0.15}_{-0.12}$ &                     $0.30^{+0.23}_{-0.15}$ \\
  3 & ELAIS-N1\_ID4 &  16:12:30.19 &  +55:58:52.27 & $3.4067\pm0.0003$ &       $\checkmark$ &                                        $3.7\pm1.0$ & $10.2^{+0.2}_{-0.3}$ &                                   $46^{+22}_{-13}$ &                       $1.16^{+0.19}_{-0.17}$ &                     $0.29^{+0.26}_{-0.18}$ \\
  4 & ELAIS-N1\_ID4 &  16:12:27.85 &  +55:57:49.74 & $3.4078\pm0.0001$ &                    &                                        $1.6\pm0.3$ &  $9.5^{+0.3}_{-0.4}$ &                                      $9^{+7}_{-4}$ &                       $0.66^{+0.27}_{-0.25}$ &                     $0.34^{+0.19}_{-0.22}$ \\
  5 & ELAIS-N1\_ID4 &  16:12:28.68 &  +56:01:27.33 & $3.4096\pm0.0001$ &       $\checkmark$ &                                        $7.8\pm0.3$ & $10.2^{+0.2}_{-0.3}$ &                                    $55^{+12}_{-9}$ &                       $0.92^{+0.10}_{-0.08}$ &                     $0.25^{+0.23}_{-0.15}$ \\
  6 & ELAIS-N1\_ID4 &  16:12:30.83 &  +55:57:49.54 & $3.4234\pm0.0002$ &                    &                                        $1.3\pm0.3$ & $10.2^{+0.2}_{-0.3}$ &                                   $51^{+15}_{-12}$ &                       $1.03^{+0.12}_{-0.12}$ &                     $0.27^{+0.22}_{-0.15}$ \\
  7 & ELAIS-N1\_ID4 &  16:12:21.83 &  +55:59:56.97 & $3.5570\pm0.0001$ &                    &                                        $5.8\pm0.4$ & $10.1^{+0.2}_{-0.3}$ &                                   $45^{+20}_{-13}$ &                       $0.98^{+0.18}_{-0.17}$ &                     $0.26^{+0.22}_{-0.15}$ \\
  8 & ELAIS-N1\_ID4 &  16:12:21.35 &  +55:57:13.52 & $3.5742\pm0.0002$ &                    &                                        $3.1\pm0.6$ &  $9.9^{+0.3}_{-0.4}$ &                                   $35^{+20}_{-12}$ &                       $1.09^{+0.24}_{-0.20}$ &                     $0.21^{+0.24}_{-0.14}$ \\
  9 & ELAIS-N1\_ID4 &  16:12:20.93 &  +55:59:12.47 & $3.6424\pm0.0002$ &                    &                                        $2.3\pm0.4$ &  $9.9^{+0.3}_{-0.4}$ &                                    $23^{+21}_{-9}$ &                       $0.95^{+0.28}_{-0.24}$ &                     $0.30^{+0.19}_{-0.20}$ \\
 10 & ELAIS-N1\_ID4 &  16:12:20.83 &  +55:58:17.74 & $3.6435\pm0.0002$ &       $\checkmark$ &                                        $6.8\pm0.7$ &  $9.7^{+0.2}_{-0.3}$ &                                     $15^{+6}_{-4}$ &                       $0.52^{+0.16}_{-0.16}$ &                     $0.32^{+0.19}_{-0.19}$ \\
 11 & ELAIS-N1\_ID4 &  16:12:32.21 &  +55:58:51.90 & $3.6476\pm0.0003$ &                    &                                        $4.0\pm0.6$ & $10.2^{+0.3}_{-0.3}$ &                                  $102^{+40}_{-31}$ &                       $1.21^{+0.14}_{-0.16}$ &                     $0.11^{+0.15}_{-0.07}$ \\
 12 &  DEEP2-3\_ID1 &  23:27:19.69 &  -00:04:07.66 & $0.1617\pm0.0001$ &                $-$ &                                                $-$ &                  $-$ &                                                $-$ &                                          $-$ &                                        $-$ \\
 13 &  DEEP2-3\_ID1 &  23:27:16.98 &  -00:03:24.52 & $0.2085\pm0.0002$ &                $-$ &                                                $-$ &                  $-$ &                                                $-$ &                                          $-$ &                                        $-$ \\
 14 &  DEEP2-3\_ID1 &  23:27:26.65 &  -00:01:52.12 & $0.2768\pm0.0003$ &                $-$ &                                                $-$ &                  $-$ &                                                $-$ &                                          $-$ &                                        $-$ \\
 15 &  DEEP2-3\_ID1 &  23:27:17.93 &  -00:04:35.55 & $3.3292\pm0.0001$ &       $\checkmark$ &                                        $7.8\pm0.4$ & $10.3^{+0.2}_{-0.2}$ &                                  $143^{+13}_{-14}$ &                       $1.00^{+0.06}_{-0.06}$ &                     $0.11^{+0.08}_{-0.05}$ \\
 16 &  DEEP2-3\_ID1 &  23:27:28.48 &  -00:04:23.25 & $3.3724\pm0.0004$ &                    &                                        $0.9\pm0.3$ &  $9.9^{+0.3}_{-0.3}$ &                                   $23^{+24}_{-10}$ &                       $1.07^{+0.35}_{-0.27}$ &                     $0.30^{+0.21}_{-0.17}$ \\
 17 &  DEEP2-3\_ID1 &  23:27:30.37 &  -00:04:48.20 & $3.4066\pm0.0001$ &                    &                                        $4.6\pm0.3$ &  $9.5^{+0.3}_{-0.3}$ &                                     $11^{+8}_{-4}$ &                       $0.60^{+0.24}_{-0.22}$ &                     $0.31^{+0.22}_{-0.19}$ \\
 18 &  DEEP2-3\_ID1 &  23:27:34.17 &  -00:04:11.29 & $3.4209\pm0.0001$ &       $\checkmark$ &                                        $4.4\pm0.4$ &  $9.9^{+0.2}_{-0.3}$ &                                    $22^{+13}_{-8}$ &                       $0.82^{+0.20}_{-0.20}$ &                     $0.31^{+0.22}_{-0.17}$ \\
 19 &  DEEP2-3\_ID1 &  23:27:17.39 &  -00:05:17.68 & $3.4279\pm0.0004$ &                    &                                        $1.9\pm0.5$ & $10.6^{+0.1}_{-0.1}$ &                                  $407^{+25}_{-27}$ &                       $1.27^{+0.03}_{-0.03}$ &                     $0.05^{+0.01}_{-0.01}$ \\
 20 &  DEEP2-3\_ID1 &  23:27:32.81 &  -00:02:14.52 & $3.4533\pm0.0002$ &       $\checkmark$ &                                        $1.6\pm0.3$ & $10.3^{+0.2}_{-0.3}$ &                                   $60^{+42}_{-22}$ &                       $1.41^{+0.25}_{-0.22}$ &                     $0.29^{+0.23}_{-0.18}$ \\
 21 &  DEEP2-3\_ID1 &  23:27:15.89 &  -00:03:02.97 & $3.4945\pm0.0001$ &       $\checkmark$ &                                        $4.0\pm0.3$ & $10.2^{+0.3}_{-0.3}$ &                                   $71^{+29}_{-21}$ &                       $1.16^{+0.17}_{-0.15}$ &                     $0.17^{+0.21}_{-0.10}$ \\
 22 &  DEEP2-3\_ID1 &  23:27:32.56 &  -00:02:37.02 & $3.5254\pm0.0001$ &                    &                                        $3.7\pm0.4$ &  $9.9^{+0.2}_{-0.3}$ &                                     $25^{+9}_{-7}$ &                       $0.68^{+0.15}_{-0.18}$ &                     $0.28^{+0.21}_{-0.19}$ \\
 23 &  DEEP2-3\_ID1 &  23:27:34.37 &  -00:02:47.45 & $3.5257\pm0.0001$ &                    &                                        $1.8\pm0.2$ & $10.1^{+0.2}_{-0.4}$ &                                   $66^{+28}_{-26}$ &                       $1.24^{+0.16}_{-0.23}$ &                     $0.15^{+0.19}_{-0.11}$ \\
 24 &  DEEP2-3\_ID1 &  23:27:17.88 &  -00:04:25.14 & $3.5810\pm0.0022$ &                    &                                        $4.7\pm1.2$ & $10.1^{+0.2}_{-0.3}$ &                                   $47^{+27}_{-17}$ &                       $1.11^{+0.22}_{-0.24}$ &                     $0.27^{+0.21}_{-0.17}$ \\
 25 &  DEEP2-3\_ID1 &  23:27:30.32 &  -00:02:24.30 & $3.5818\pm0.0005$ &                    &                                        $1.3\pm0.4$ & $10.0^{+0.2}_{-0.2}$ &                                  $104^{+51}_{-44}$ &                       $1.45^{+0.13}_{-0.17}$ &                     $0.02^{+0.03}_{-0.01}$ \\
 26 &  DEEP2-3\_ID1 &  23:27:25.00 &  -00:03:17.58 & $3.6090\pm0.0002$ &                    &                                        $1.6\pm0.4$ &  $9.9^{+0.3}_{-0.3}$ &                                   $49^{+23}_{-18}$ &                       $1.14^{+0.17}_{-0.22}$ &                     $0.11^{+0.19}_{-0.08}$ \\
 27 &  DEEP2-3\_ID1 &  23:27:16.82 &  -00:02:09.09 & $3.6361\pm0.0033$ &                    &                                        $1.6\pm0.5$ & $10.0^{+0.3}_{-0.5}$ &                                   $47^{+28}_{-17}$ &                       $1.21^{+0.22}_{-0.20}$ &                     $0.14^{+0.23}_{-0.11}$ \\
 28 &  DEEP2-3\_ID1 &  23:27:23.14 &  -00:03:11.99 & $3.6424\pm0.0001$ &       $\checkmark$ &                                        $5.6\pm0.4$ & $10.0^{+0.3}_{-0.4}$ &                                   $29^{+21}_{-10}$ &                       $0.84^{+0.26}_{-0.20}$ &                     $0.34^{+0.19}_{-0.21}$ \\
 29 &  DEEP2-3\_ID2 &  23:29:44.05 &  +00:42:58.27 & $3.1289\pm0.0001$ &                    &                                        $3.9\pm0.5$ & $10.0^{+0.2}_{-0.3}$ &                                    $26^{+15}_{-6}$ &                       $0.67^{+0.25}_{-0.13}$ &                     $0.36^{+0.23}_{-0.21}$ \\
 30 &  DEEP2-3\_ID2 &  23:29:59.03 &  +00:45:21.56 & $3.2509\pm0.0001$ &                    &                                        $4.1\pm0.4$ & $10.9^{+0.3}_{-0.4}$ &                                $226^{+172}_{-103}$ &                       $1.84^{+0.26}_{-0.32}$ &                     $0.35^{+0.23}_{-0.21}$ \\
 31 &  DEEP2-3\_ID2 &  23:29:48.67 &  +00:43:15.42 & $3.3901\pm0.0001$ &       $\checkmark$ &                                        $7.3\pm0.2$ & $10.5^{+0.2}_{-0.3}$ &                                   $93^{+31}_{-24}$ &                       $1.18^{+0.15}_{-0.14}$ &                     $0.33^{+0.21}_{-0.19}$ \\
 32 &  DEEP2-3\_ID2 &  23:29:58.29 &  +00:42:39.88 & $3.3904\pm0.0002$ &                    &                                        $3.7\pm0.4$ & $10.4^{+0.2}_{-0.2}$ &                                   $71^{+13}_{-10}$ &                       $0.73^{+0.09}_{-0.07}$ &                     $0.30^{+0.23}_{-0.15}$ \\
 33 &  DEEP2-3\_ID2 &  23:29:51.99 &  +00:42:30.12 & $3.4027\pm0.0001$ &                    &                                        $2.9\pm0.4$ & $10.1^{+0.3}_{-0.3}$ &                                   $41^{+19}_{-12}$ &                       $0.86^{+0.18}_{-0.15}$ &                     $0.32^{+0.22}_{-0.20}$ \\
 34 &  DEEP2-3\_ID2 &  23:29:57.40 &  +00:43:04.30 & $3.4032\pm0.0003$ &                    &                                        $2.6\pm0.5$ & $10.8^{+0.3}_{-0.3}$ &                                 $175^{+102}_{-55}$ &                       $1.56^{+0.21}_{-0.17}$ &                     $0.33^{+0.20}_{-0.18}$ \\
 35 &  DEEP2-3\_ID2 &  23:29:55.85 &  +00:44:17.32 & $3.4293\pm0.0001$ &                    &                                        $2.4\pm0.3$ & $10.2^{+0.4}_{-0.4}$ &                                   $50^{+55}_{-28}$ &                       $1.33^{+0.35}_{-0.37}$ &                     $0.32^{+0.22}_{-0.21}$ \\
 36 &  DEEP2-3\_ID2 &  23:29:49.42 &  +00:42:56.47 & $3.5265\pm0.0001$ &                    &                                        $4.8\pm0.3$ & $10.1^{+0.2}_{-0.3}$ &                                   $35^{+17}_{-13}$ &                       $0.78^{+0.19}_{-0.21}$ &                     $0.37^{+0.17}_{-0.19}$ \\
 37 &  DEEP2-3\_ID2 &  23:29:57.41 &  +00:42:31.91 & $3.6213\pm0.0001$ &                    &                                        $4.4\pm0.3$ &  $9.7^{+0.3}_{-0.4}$ &                                     $14^{+7}_{-5}$ &                       $0.50^{+0.19}_{-0.20}$ &                     $0.34^{+0.17}_{-0.20}$ \\
 38 &  DEEP2-3\_ID2 &  23:29:57.37 &  +00:43:56.88 & $3.7284\pm0.0002$ &                    &                                        $0.9\pm0.3$ & $10.0^{+0.4}_{-0.5}$ &                                   $36^{+34}_{-17}$ &                       $1.05^{+0.27}_{-0.26}$ &                     $0.28^{+0.21}_{-0.20}$ \\
 39 &  DEEP2-3\_ID2 &  23:29:56.42 &  +00:42:14.34 & $3.7824\pm0.0002$ &                    &                                        $1.1\pm0.3$ & $10.1^{+0.4}_{-0.4}$ &                                   $41^{+38}_{-21}$ &                       $1.03^{+0.30}_{-0.29}$ &                     $0.29^{+0.18}_{-0.19}$ \\
 40 &  DEEP2-3\_ID2 &  23:29:56.18 &  +00:44:47.25 & $3.9471\pm0.0002$ &                    &                                        $5.9\pm0.8$ &  $9.9^{+0.3}_{-0.4}$ &                                   $30^{+18}_{-12}$ &                       $0.70^{+0.21}_{-0.21}$ &                     $0.23^{+0.20}_{-0.15}$ \\
 41 &   COSMOS\_ID0 &  09:57:32.17 &  +01:00:35.30 & $3.2430\pm0.0001$ &       $\checkmark$ &                                        $4.9\pm0.3$ & $10.5^{+0.2}_{-0.2}$ &                                  $154^{+25}_{-26}$ &                       $1.36^{+0.09}_{-0.09}$ &                     $0.17^{+0.13}_{-0.09}$ \\
 42 &   COSMOS\_ID0 &  09:57:26.64 &  +01:00:13.49 & $3.2811\pm0.0004$ &                    &                                        $1.1\pm0.2$ & $10.3^{+0.2}_{-0.3}$ &                                  $129^{+27}_{-25}$ &                       $1.38^{+0.10}_{-0.10}$ &                     $0.12^{+0.15}_{-0.06}$ \\
 43 &   COSMOS\_ID0 &  09:57:31.11 &  +00:56:21.29 & $3.3151\pm0.0001$ &       $\checkmark$ &                                        $3.4\pm0.2$ &  $9.9^{+0.2}_{-0.3}$ &                                     $23^{+8}_{-6}$ &                       $0.77^{+0.15}_{-0.12}$ &                     $0.31^{+0.22}_{-0.20}$ \\
 44 &   COSMOS\_ID0 &  09:57:38.69 &  +00:59:28.17 & $3.3165\pm0.0003$ &                    &                                        $0.7\pm0.2$ & $10.1^{+0.3}_{-0.4}$ &                                   $34^{+21}_{-14}$ &                       $1.10^{+0.22}_{-0.27}$ &                     $0.33^{+0.22}_{-0.21}$ \\
 45 &   COSMOS\_ID0 &  09:57:36.92 &  +00:58:43.07 & $3.3168\pm0.0001$ &                    &                                        $3.9\pm0.3$ & $10.5^{+0.3}_{-0.2}$ &                                  $195^{+38}_{-35}$ &                       $1.49^{+0.11}_{-0.10}$ &                     $0.12^{+0.17}_{-0.07}$ \\
\bottomrule
\end{tabular}

%% file: tbl/oiii_detected_galaxies_part2.tex
\begin{tabular}{rcccccccccc}
\toprule
 ID &      Region & R.A. (J2000) & Decl. (J2000) & $z_\mathrm{spec}$ & H$\beta$ Detection & \begin{tabular}{c}[O\textsc{iii}]5007 Flux\\ (10$^{-17}$ erg/s/cm$^2$)\end{tabular} &  $\log{M_*/M_\odot}$ & \begin{tabular}{c}SFR\\ (M$_\odot$/yr)\end{tabular} & \begin{tabular}{c}$A_V$\\ (mag)\end{tabular} & \begin{tabular}{c}Age\\ (Gyr)\end{tabular} \\
\midrule
 46 & COSMOS\_ID0 &  09:57:34.61 &  +00:57:44.67 & $3.3175\pm0.0001$ &       $\checkmark$ &                                        $2.3\pm0.2$ & $10.5^{+0.2}_{-0.3}$ &                                  $143^{+36}_{-33}$ &                       $1.46^{+0.11}_{-0.13}$ &                     $0.18^{+0.22}_{-0.11}$ \\
 47 & COSMOS\_ID0 &  09:57:26.24 &  +01:01:37.54 & $3.4078\pm0.0002$ &       $\checkmark$ &                                        $1.5\pm0.2$ & $10.2^{+0.2}_{-0.2}$ &                                   $90^{+18}_{-17}$ &                       $1.15^{+0.09}_{-0.11}$ &                     $0.12^{+0.15}_{-0.06}$ \\
 48 & COSMOS\_ID0 &  09:57:26.15 &  +00:58:41.83 & $3.4213\pm0.0001$ &                    &                                        $4.2\pm0.3$ &  $9.9^{+0.2}_{-0.3}$ &                                     $23^{+7}_{-6}$ &                       $0.71^{+0.13}_{-0.13}$ &                     $0.27^{+0.24}_{-0.15}$ \\
 49 & COSMOS\_ID0 &  09:57:22.75 &  +00:57:02.95 & $3.4236\pm0.0003$ &                    &                                        $0.6\pm0.2$ & $10.1^{+0.3}_{-0.3}$ &                                   $36^{+32}_{-17}$ &                       $1.26^{+0.30}_{-0.30}$ &                     $0.31^{+0.21}_{-0.17}$ \\
 50 & COSMOS\_ID0 &  09:57:29.80 &  +00:58:15.16 & $3.4245\pm0.0003$ &                    &                                        $1.9\pm0.2$ & $10.2^{+0.2}_{-0.3}$ &                                   $55^{+19}_{-14}$ &                       $1.06^{+0.14}_{-0.13}$ &                     $0.25^{+0.22}_{-0.15}$ \\
 51 & COSMOS\_ID0 &  09:57:32.08 &  +00:58:28.63 & $3.4258\pm0.0001$ &                    &                                        $0.9\pm0.2$ &  $9.6^{+0.2}_{-0.3}$ &                                     $11^{+6}_{-3}$ &                       $0.51^{+0.20}_{-0.15}$ &                     $0.29^{+0.22}_{-0.18}$ \\
 52 & COSMOS\_ID0 &  09:57:28.74 &  +00:59:41.17 & $3.4275\pm0.0001$ &       $\checkmark$ &                                        $7.3\pm0.2$ & $10.2^{+0.2}_{-0.2}$ &                                   $67^{+15}_{-11}$ &                       $0.93^{+0.10}_{-0.09}$ &                     $0.20^{+0.19}_{-0.11}$ \\
 53 & COSMOS\_ID0 &  09:57:27.06 &  +00:58:53.63 & $3.4277\pm0.0003$ &                    &                                        $1.6\pm0.3$ & $10.0^{+0.2}_{-0.2}$ &                                     $55^{+8}_{-9}$ &                       $0.77^{+0.08}_{-0.08}$ &                     $0.17^{+0.16}_{-0.09}$ \\
 54 & COSMOS\_ID0 &  09:57:27.11 &  +00:57:39.36 & $3.4351\pm0.0001$ &                    &                                        $2.0\pm0.2$ &  $9.5^{+0.3}_{-0.3}$ &                                     $10^{+8}_{-4}$ &                       $0.54^{+0.28}_{-0.19}$ &                     $0.29^{+0.23}_{-0.18}$ \\
 55 & COSMOS\_ID0 &  09:57:35.04 &  +01:01:11.15 & $3.5444\pm0.0002$ &                    &                                        $1.2\pm0.2$ & $10.3^{+0.3}_{-0.3}$ &                                  $126^{+61}_{-35}$ &                       $1.36^{+0.20}_{-0.19}$ &                     $0.11^{+0.17}_{-0.08}$ \\
 56 & COSMOS\_ID0 &  09:57:25.96 &  +01:00:54.33 & $3.5810\pm0.0001$ &                    &                                        $1.3\pm0.2$ &  $9.1^{+0.3}_{-0.3}$ &                                      $4^{+3}_{-1}$ &                       $0.25^{+0.24}_{-0.17}$ &                     $0.33^{+0.18}_{-0.19}$ \\
 57 & COSMOS\_ID0 &  09:57:34.96 &  +00:58:25.65 & $3.6567\pm0.0001$ &                    &                                        $2.1\pm0.3$ &  $9.8^{+0.2}_{-0.3}$ &                                    $18^{+10}_{-6}$ &                       $0.65^{+0.20}_{-0.21}$ &                     $0.37^{+0.17}_{-0.20}$ \\
 58 & COSMOS\_ID0 &  09:57:34.08 &  +00:59:18.16 & $3.6940\pm0.0001$ &                    &                                        $3.9\pm0.3$ & $10.1^{+0.2}_{-0.3}$ &                                   $46^{+15}_{-10}$ &                       $0.80^{+0.14}_{-0.11}$ &                     $0.23^{+0.21}_{-0.14}$ \\
 59 & COSMOS\_ID0 &  09:57:33.85 &  +00:56:44.18 & $3.6952\pm0.0001$ &       $\checkmark$ &                                       $17.1\pm0.4$ & $10.4^{+0.2}_{-0.3}$ &                                  $107^{+12}_{-12}$ &                       $0.69^{+0.09}_{-0.06}$ &                     $0.22^{+0.20}_{-0.14}$ \\
 60 & COSMOS\_ID0 &  09:57:33.03 &  +01:00:40.31 & $3.7361\pm0.0001$ &       $\checkmark$ &                                        $8.3\pm0.4$ & $10.1^{+0.3}_{-0.2}$ &                                    $64^{+10}_{-7}$ &                       $0.66^{+0.08}_{-0.07}$ &                     $0.14^{+0.17}_{-0.07}$ \\
\bottomrule
\end{tabular}